\documentclass[aps,prb,preprint]{revtex4}
\usepackage{subfigure}
\usepackage{graphicx}
\usepackage{dcolumn}
\usepackage{bm}
\usepackage{textcomp}
\usepackage{amsmath}
\usepackage{amssymb}
\usepackage{epstopdf}
\usepackage{array}
\usepackage[normalem]{ulem}
\graphicspath{{./FIGS/}}

\newcommand{\kv}{{\bf k}}

\newcommand{\ep}{\epsilon}


\preprint{}
\begin{document}

\title{Mexican Hat and Rashba Bands in Few-Layer van der Waals Materials}

\author{Darshana Wickramaratne}
\affiliation{LAboratory for Terascale and Terahertz Electronics, Department of Electrical
and Computer Engineering,
University of California, Riverside, CA 92521}

\author{Ferdows Zahid}
\affiliation{Department of Physics and the Center of Theoretical and Computational Physics,
The University of Hong Kong, Pokfulam Road, Hong Kong SAR, China}

\author{Roger K. Lake}
\affiliation{LAboratory for Terascale and Terahertz Electronics, Department of Electrical
and Computer Engineering,
University of California, Riverside, CA 92521}

\begin{abstract}
The valence band of a variety of few-layer, two-dimensional materials
consists of a ring of states in the Brillouin zone.
The energy-momentum relation has the form of a `Mexican hat'
or a Rashba dispersion.
The two-dimensional density of states is singular
at or near the band edge,
and the band-edge density of modes turns on nearly abruptly as a step function.
The large band-edge density of modes 
enhances the Seebeck coefficient, the power factor, 
and the thermoelectric figure of merit ZT.
Electronic and thermoelectric 
properties are determined from ab initio calculations
for few-layer III-VI materials GaS, GaSe, InS, InSe, for Bi$_{2}$Se$_{3}$,
for monolayer Bi,
and for bilayer graphene as a function of vertical field.
The effect of interlayer coupling on these properties in few-layer III-VI materials and 
Bi$_{2}$Se$_{3}$ is described.
Analytical models provide insight into
the layer dependent trends that are relatively consistent
for all of these few-layer materials. 
Vertically biased bilayer graphene could serve as an experimental test-bed 
for measuring these effects.
\end{abstract}
\maketitle

\newpage

\section{Introduction}
The electronic bandstructure of many two-dimensional (2D), van der Waals (vdW) materials
qualitatively changes as the thickness is reduced down to a few
monolayers.
One well known example is
the indirect to direct gap transition that
occurs at monolayer thicknesses of the Mo and W 
transition metal dichalcogenides (TMDCs)\cite{MoS2_Mak_Heinz}.
Another qualitative change that occurs in 
a number of 2D materials is the inversion of the parabolic
dispersion at a band extremum into a `Mexican hat' 
dispersion.\cite{Fermi_ring_Neto_PRB07,zolyomi_GaX,zolyomi_InX}
Mexican hat dispersions are also 
referred to as a Lifshiftz 
transition \cite{graphene_trilayer_Macdonald,zolyomi_GaX,Falko_BLG_Lifshitz_PRL14},
an electronic topological transition \cite{blanter_ETT_theory} or a camel-back
dispersion \cite{PbTe_camelback_thermo,Te_camelback_thermo}.
In a Mexican hat dispersion, the Fermi surface near the 
band-edge is approximately a ring in $k$-space,
and the radius of the ring can be large, on the order of half of the
Brillouin zone.
The large degeneracy coincides with a singularity in the 
two-dimensional (2D) density of states close to the band edge.
A similar feature occurs in monolayer Bi due to the Rashba splitting
of the valence band.
This also results in a valence band edge that is a ring in $k$-space
although the diameter of the ring is generally smaller than that of the 
Mexican hat dispersion.

Mexican hat dispersions are relatively common in few-layer two-dimensional
materials.
Ab-initio studies have found
Mexican hat dispersions in the valence band of many few-layer III-VI
materials such as GaSe, GaS, InSe, InS \cite{zolyomi_GaX, zolyomi_InX, GaS_photodetector_AnPingHu,
Hennig_GroupIII_ChemMat, SGLouie_GaSe_arxiv, WYao_GaS_GaSe_arxiv}. 
Experimental studies have demonstrated synthesis of monolayers
and or few layers of GaS, GaSe and InSe thin films.\cite{GaSe_Ajayan_NL13,Ajayan_InSe,
GaS_photodetector_AnPingHu, Xiao_GaSe_nanosheets,Dravid_GaS_GaSe_AdvMat,
GaSe_Geohagen_SciRep, GaSe_Geohagen_ACSNano, CZhou_GaS_ACSNano}.
Monolayers of Bi$_2$Te$_3$ \cite{Zahid_Lake}, 
and Bi$_{2}$Se$_{3}$ \cite{Udo_Bi2Se3} also exhibit a Mexican
hat dispersion in the valence band.
The conduction and valence bands of bilayer graphene
distort into approximate Mexican hat dispersions, with considerable anisotropy, when a 
a vertical field is applied across
AB-stacked bilayer graphene. 
\cite{MacDonald_bi_gap_PRB07,Fermi_ring_Neto_PRB07,Falko_BLG_Lifshitz_PRL14}
The valence band of monolayer Bi(111) has a Rashba dispersion.
\cite{Bi_ARPES_Rashba_PRL11}

The large density of states of the Mexican hat dispersion can lead to instabilities near
the Fermi level, and two
different ab initio studies have recently predicted Fermi-level controlled
magnetism in monolayer GaSe and GaS \cite{SGLouie_GaSe_arxiv, WYao_GaS_GaSe_arxiv}.
The singularity in the density of states 
and the large number of conducting modes at the band edge 
can enhance the 
Seebeck coefficient, power factor, and the thermoelectric figure of merit ZT.
\cite{Mahan:1996,heremans_PbTe_thermo,heremans_distortionDOS_Thermo}
Prior studies have achieved this enhancement in the density of
states by using nanowires \cite{Dressel:1D:PRB:1993,Dresselhaus_New_Directions07}, 
introducing resonant doping 
levels \cite{heremans_PbTe_thermo,heremans_distortionDOS_Thermo}, 
high band degeneracy \cite{Snyder_degenerate_thermo,Kanatzidis_New_Old_AChemie09}, 
and using the Kondo resonance associated with the 
presence of localized $d$ and $f$ 
orbitals \cite{Kondo_thermo_Cava,Steglich_FeSb2_thermo,Gang_Chen_Kondo_NW11}.
The large increase in ZT predicted for monolayer Bi$_2$Te$_3$ resulted from the
formation of a Mexican hat bandstructure and its large band-edge
degeneracy \cite{Zahid_Lake,Lundstrom_Jesse_Bi2Te3}.

\begin{figure}[!h]
\includegraphics[width=5in]{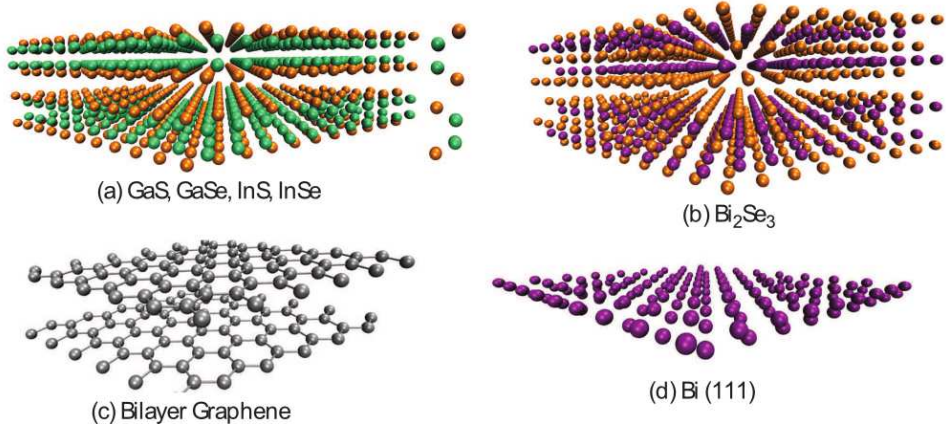}
\caption{(Color online) Atomic structures of van-der Waals materials
with a Mexican hat or Rashba dispersion: 
(a) Bilayer III-VI material. The $\beta$ phase stacking geometry is shown at right.
(b)Bi$_{2}$Se$_{3}$, (c) Bilayer Graphene and (d) Bi(111) monolayer 
}
\label{fig:structure}
\end{figure}

This work theoretically investigates the electronic and thermoelectric properties of a variety of
van der Waals materials that exhibit a Mexican hat dispersion or Rashba dispersion.
The Mexican hat and Rashba dispersions are first analyzed using an 
analytical model.
Then, density functional theory is used to calculate 
the electronic and thermoelectric properties of bulk and one 
to four monolayers 
of GaX, InX (X = Se, S), Bi$_{2}$Se$_{3}$, monolayer Bi(111),
and bilayer graphene as a function of vertical electric field.
Figure \ref{fig:structure} illustrates the 
investigated structures that have either a Mexican hat or
Rashba dispersion.
The analytical model combined with the numerically calculated orbital compositions 
of the conduction
and valence bands explain the layer dependent trends that are relatively consistent
for all of the few-layer materials.
While numerical values are provided for various thermoelectric metrics, the emphasis
is on the layer-dependent trends and the analysis of how the bandstructure
affects both the electronic and thermoelectric properties.
The metrics are provided in such a way that new estimates can be readily obtained
given new values for the electrical or thermal conductivity.

\section{Models and Methods}
\subsection{Landauer Thermoelectric Parameters}
In the linear response regime, the electronic and thermoelectric parameters
are calculated within a Landauer 
formalism.
The basic equations have been described previously \cite{Lundstrom_Jesse_Bi2Te3,Klimeck_DOM_thermoelectric_JCE,Darshana_MX2_Thermo}, and we list them below for convenience.
The equations for the 
electronic conductivity ($\sigma$), the electronic thermal conductivity ($\kappa_{e}$),
and the Seebeck coefficient (S) are
\begin{align}
\sigma &= (2q^{2}/h)I_{0}\quad (\mathrm{\Omega^{-1} m^{2-D}}),
\label{eq:sigma}\\
\kappa_{e} &= (2Tk_{B}^{2}/h)(I_{2} - I_{1}^{2}/I_{0}) \quad (\mathrm{W  m^{2-D} K^{-1}}),
\label{eq:Ke}\\
S &= -(k_{B}/q)\frac{I_{1}}{I_{0}}\quad (\mathrm{V/K}),
\label{eq:S}\\
\mathrm{with} \nonumber \\
I_{j} &= L \int_{-\infty}^{\infty} \left(\frac{E-E_{F}}{k_{B}T}\right)^{j} 
\bar{T}(E)
\left(-\frac{\partial f}{\partial E}\right)dE  
\label{eq:Ij}
\end{align}
where $L$ is the device length, $D$ is the dimensionality (1, 2, or 3),
$q$ is the magnitude of the electron charge, $h$
is Planck's constant, $k_B$ is Boltzmann's constant,
and $f$ is the Fermi-Dirac factor. 
The transmission function $\bar{T}$ is
\begin{equation}
\bar{T}(E) = T(E)M(E)
\label{eq:TE}
\end{equation}
where M(E) as the density of modes. 
In the diffusive limit,
\begin{equation}
T(E)=\lambda(E)/L , 
\label{eq:TE_diff}
\end{equation}
where $\lambda(E)$ is the electron mean free path. 
The power factor ($PF$) and the thermoelectric figure of merit ($ZT$)
are given by
$PF = S^2 \sigma$ and 
\begin{equation}
ZT = S^2 \sigma T / (\kappa_{l} + \kappa_{e})
\label{eq:ZT}
\end{equation}
where $\kappa_l$ is the lattice thermal conductivity.

\subsection{Analytical Models}
\label{sec:analytical}

The single-spin density of modes for transport in the $x$ direction 
is \cite{Datta_book05,Lundstrom:TE:TB:JAP:2010} 
\begin{equation}
M(E)= \frac{2\pi}{L^D} \sum_{\kv} \delta(E - \epsilon(\kv)) \frac{\partial \ep}{\partial k_x}
\label{eq.ME}
\end{equation}
where $D$ is the dimensionality, $E$ is the energy, and $\ep(\kv)$ is the band dispersion.
The sum is over all values of $\kv$ such that $\frac{\partial \ep}{\partial k_x} > 0$,
i.e. all momenta with positive velocities.
The dimensions are $1/L^{D-1}$, so that in 2D, $M(E)$ gives the number of 
modes per unit width at energy $E$.
If the dispersion is only a function of the magnitude of $\kv$, then
Eq. (\ref{eq.ME}) reduces to
\begin{equation}
M(E)= \frac{N_D}{\left(2\pi\right)^{D-1}} \sum_{b} k_b^{D-1}(E)
\label{eq.MEk}
\end{equation}
where $N_D = \pi$ for $D=3$, $N_D = 2$ for $D=2$, and $N_D = 1$ for $D=1$.
$k_b$ is the magnitude of $\kv$ such that $E = \ep(k_b)$, 
and the sum is over all bands and all values of $k_b$ within a band.
When a band-edge is a ring in $k$-space with radius $k_0$, the 
single-spin 2D density of modes at the band edge is 
\begin{equation}
M(E_{\rm edge}) = N \frac{k_0}{\pi} ,
\label{eq:M_edge}
\end{equation}
where $N$ is either 1 or 2 depending on the type of dispersion, Rashba or 
Mexican hat.
Thus, the 2D density of modes at the band edge depends only on the radius of
of the $k$-space ring. 
%
%
For a two dimensional parabolic dispersion, $E = \frac{\hbar^2k^2}{2m^*}$,
the radius is 0, and
Eq. (\ref{eq.MEk}) gives a
the single-spin density of modes of \cite{Lundstrom:TE:analytic:JAP:2009}
\begin{equation}
M_{\rm par}(E)=\frac{\sqrt{2m^*E}}{\pi\hbar}. 
\label{eq:DOM_2D}
\end{equation}

\begin{figure}[!h]
\includegraphics[width=5in]{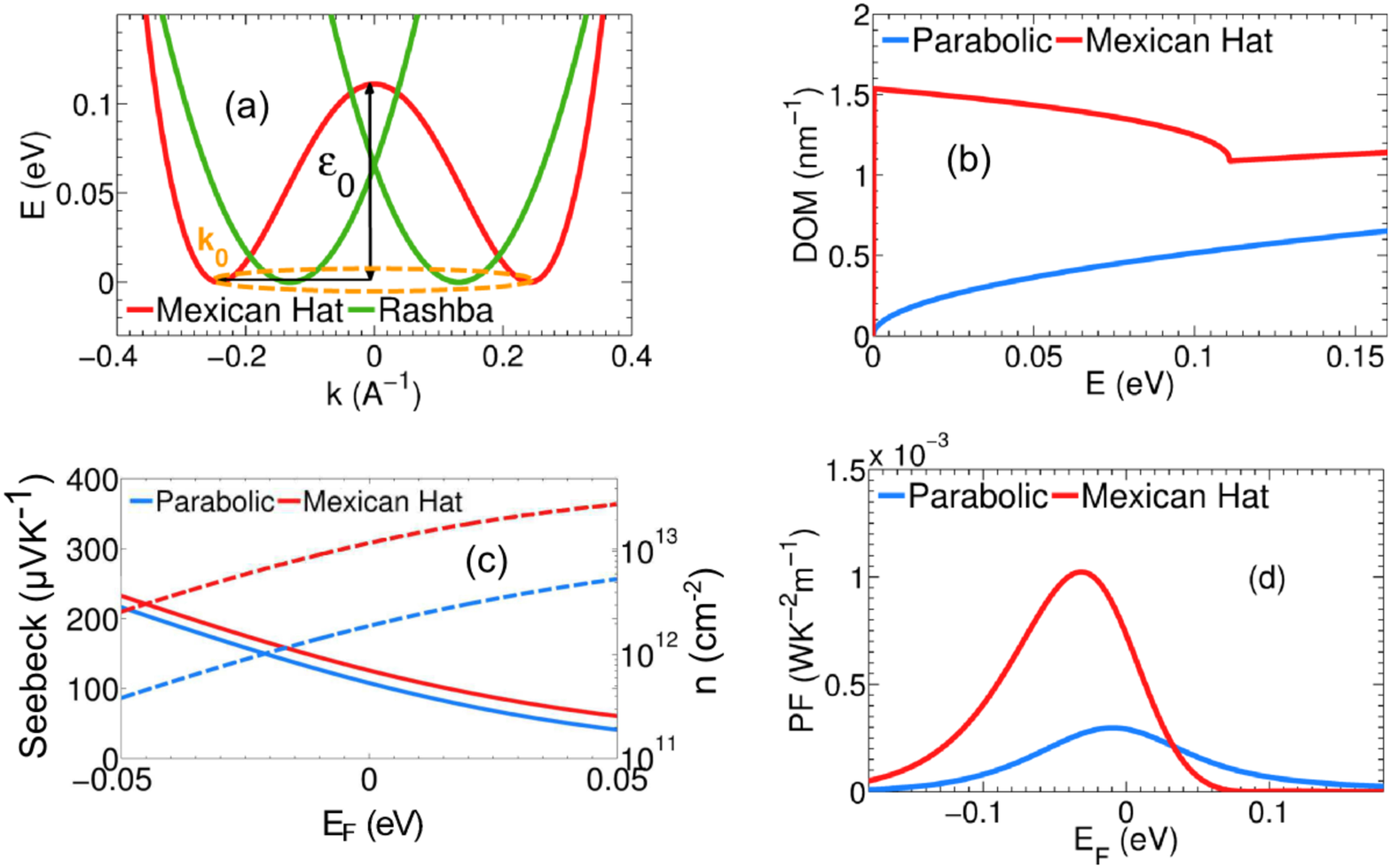}
\caption{(Color online) 
(a)  Comparison of a Mexican hat dispersion (red) and a Rashba dispersion (green).  
The band edges are rings in $k$-space with radius $k_0$ 
illustrated for the Mexican hat band by the orange dotted circle.  
The height of the Mexican hat band at $k=0$ is $\epsilon_{0} = 0.111$ eV.
The Rashba parameter is 1.0 eV \AA, and the effective mass for both dispersions
is the bare electron mass $m_0$.
(b) Density of modes of the Mexican hat dispersion (red) versus 
parabolic band (blue).
The parabolic dispersion also has an effective mass of 1.0.
(c) Room temperature Seebeck coefficients (solid lines) and carrier concentrations (broken lines)
of the Mexican hat band (red) and the parabolic band (blue) as a function of Fermi level
position, E$_{F}$.
(d) Room temperature ballistic power factor of the Mexican hat band (red) and the 
parabolic band (blue) calculated from Eqs. (\ref{eq:sigma}), 
(\ref{eq:S}), and (\ref{eq:Ij}) with $T(E) = 1$.
}
\label{fig:MexicanHat}
\end{figure}

In real III-VI materials, there is anisotropy in 
the Fermi surfaces, and a 6th order, angular dependent polynomial expression 
is provided by Z\'{o}lyomi et al. that captures the low-energy
anisotropy \cite{zolyomi_GaX,zolyomi_InX}.
To obtain physical insight with closed form expressions, 
we consider a 4th order analytical form for an isotropic Mexican hat dispersion 
\begin{equation}
\epsilon(k) = \epsilon_{0} - \frac{\hbar^{2} k^2}{2m^*}
+ \frac{1}{4\epsilon_{0}} \left( \frac{\hbar^{2} k^2}{2m^*} \right)^2 
\label{eq:MexHat_Ek}
\end{equation}
where $\ep_0$ is the height of the hat at $k=0$,
and $m^*$ is the magnitude of the
effective mass at $k=0$.
A similar quartic form was previously used to analyze the 
effect of electron-electron interactions in
biased bilayer graphene \cite{Fermi_ring_Neto_PRB07}.
The function is plotted in Figure \ref{fig:MexicanHat}(a). 
The band edge occurs at $\ep = 0$, and, in $k$-space,
in two dimensions (2D), it forms a ring in the $k_x-k_y$ plane
with a radius of 
\begin{equation}
k_0^{\rm MH} = \frac{2 \: \sqrt{m^* \ep_0}}{\hbar} . 
\label{eq:k0_MHat}
\end{equation}
%
%
For the two-dimensional Mexican hat dispersion of Eq. (\ref{eq:MexHat_Ek}),
the single-spin density of modes is
\begin{equation}
M_{\rm MH}(E)=
\left\{
\begin{array}{ll}
\frac{k_0^{\rm MH}}{\pi}
\left( \sqrt{1+\sqrt\frac{E}{\epsilon_{0}}} + \sqrt{1-\sqrt\frac{E}{\epsilon_{0}}} \; \right) 
\;\;\;\;\;
&
\left( 0 \leq E \leq \ep_0 \right)
\\ 
\frac{k_0^{\rm MH}}{\pi}
\left( \sqrt{1+\sqrt\frac{E}{\epsilon_{0}}} \; \right) 
&
\left( \ep_0 \leq E \right) .
\end{array}
\right.
\label{eq.ME_hat}
\end{equation}
Figure \ref{fig:MexicanHat}(b) shows the density of mode distributions 
plotted from Eqs. (\ref{eq:DOM_2D}) and (\ref{eq.ME_hat}).
At the band edge ($E=0$), the single-spin density of modes 
of the Mexican hat dispersion is finite,
\begin{equation}
M_{\rm MH}(E=0) = \frac{2 k_0^{\rm MH}}{\pi} .
\label{eq.ME_bandedge}
\end{equation}
The Mexican hat density of modes decreases by a factor of $\sqrt{2}$
as the energy increases from 0 to $\ep_0$, and then it slowly increases. 
%
The step-function turn-on of the density of modes is associated with
a singularity in the density of states. 
The single-spin density of states resulting from the Mexican hat dispersion is
\begin{equation}
D_{\rm MH}(E)=
\left\{
\begin{array}{ll}
\frac{m^*}{\pi \hbar^2} \sqrt{ \frac{\ep_0}{E} } \;\;\;\;
&
(0 \leq E \leq \ep_0)\\ 
\frac{m^*}{2 \pi \hbar^2 } \sqrt{ \frac{\ep_0}{E} }
&
\left( \ep_0 < E \right) \; .
\end{array}
\right. 
\label{eq.DE_hat}
\end{equation}
%
%

Rashba splitting of the spins also results in a valence band edge that is
a ring in $k$-space.
The Bychkov-Rashba
model with linear and quadratic terms in $k$ gives 
an analytical expression for a 
Rashba-split dispersion, \cite{Bychkov_Rashba_JPhysC}
\begin{equation}
\ep(\kv) = \ep_0 + \frac{h^{2}k^{2}}{2m^*} \pm \alpha_{R} k 
\label{eq:MexHat_Rashba}
\end{equation}
where the Rashba parameter, $\alpha_{R}$, 
is the strength of the Rashba splitting.
In Eq. (\ref{eq:MexHat_Rashba}), the bands are shifted up by 
$\ep_0 = \frac{\alpha_R^2 m^*}{2 \hbar^2}$ 
so that the band edge occurs at $\ep = 0$. 
The radius of the band edge in $k$-space is 
\begin{equation}
k_0^{\rm R} = \frac{m^* \alpha_R}{\hbar^2} = \frac{\sqrt{2 m^* \ep_0}}{\hbar}.
\label{eq:k0_Rashba}
\end{equation}
The energy dispersion of the split bands is illustrated in 
Figure \ref{fig:MexicanHat}(a).
The density of modes, {\em including both spins}, 
resulting from the dispersion of Eq. (\ref{eq:MexHat_Rashba}) is
\begin{equation}
M_{\rm R}^{\rm 2 \: spins}(E) = 
\left\{
\begin{array}{ll}
\frac{2 k_0^{\rm R}}{\pi}
&
\;\;\;\;\; \left( 0 \leq E \leq \ep_0 \right)\\
\frac{2 k_0^{\rm R}}{\pi} \sqrt{ \frac{E}{\ep_0} } 
&
\;\;\;\;\; \left( \ep_0 \leq E \right)
\end{array}
\right.
\label{eq:MexHat_DOM}
\end{equation}
For $0 \leq E \leq \ep_0$, the density of modes is a step function
and the height is determined by $\alpha_R$ and the effective mass.
Values for 
$\alpha_{R}$ vary from 0.07 eV\AA~in InGaAs/InAlAs quantum
wells to 0.5 eV\AA~in the Bi(111) monolayer.\cite{BiTeI_Rashba_NatMat11}
The density of states including both spins is
\begin{equation}
D_{\rm R}(E)=
\left\{
\begin{array}{ll}
\frac{m^*}{\pi \hbar^2} \sqrt{ \frac{\ep_0}{E} } \;\;\;\;
&
(0 \leq E \leq \ep_0)\\ 
\frac{m^*}{\pi \hbar^2 } 
&
\left( \ep_0 < E \right)
\end{array}
\right.
\label{eq.DE_R}
\end{equation}
In general, we find that the diameter of the Rashba $k$-space rings
are less than the diameter of the Mexican hat
$k$-space rings, so that the enhacements to the thermoelectric
parameters are less from Rashba-split bands than from the inverted
Mexican hat bands.

In the real bandstructures considered in the Sec. \ref{sec:num_results}, 
there is anisotropy to the $k$-space Fermi surfaces.
The band extrema at K and M have different energies. 
For the III-VIs, Bi$_2$Se$_3$, and monolayer Bi, 
this energy difference is less than $k_BT$ at room temperature.
In the III-VIs, the maximum energy difference between the valence band
extrema at K and M is 6.6 meV in InS.
In Bi$_2$Se$_3$, it is 19.2 meV, and in monolayer Bi, it is 0.5 meV.
The largest anisotropy occurs in bilayer graphene under bias. 
At the maximum electric field considered of 0.5 V/\AA, the energy
difference of the extrema in the conduction band is 112 meV, and the 
energy difference of the extrema in the valence band is 69 meV.
The anisotropy experimentally manifests itself in the 
quantum Hall plateaus.\cite{Falko_BLG_Lifshitz_PRL14}
Anisotropy results in a finite slope to the turn-on of the density of modes and
a shift of the singularity in the density of states away from the band edge.
The energy of the singularity in the density of states 
lies between the two extrema \cite{zolyomi_GaX,zolyomi_InX}.

Figure \ref{fig:MexicanHat}(c) compares the Seebeck coefficients
and the electron densities calculated from the
Mexican hat dispersion shown in Fig. \ref{fig:MexicanHat}(a)
and a parabolic dispersion.
The quantities are plotted versus Fermi energy with the conduction
band edge at $E=0$.
The bare electron mass is used for both dispersions, $m^* = m_0$,
and, for the Mexican hat,
$\epsilon_{0} = 0.111$ eV which is the largest value 
for $\epsilon_{0}$ obtained from our ab-initio simulations of the III-VI compounds.
The temperature is $T = 300$ K.
The Seebeck coefficients are calculated from Eqs. 
(\ref{eq:S}), (\ref{eq:Ij}), and (\ref{eq:TE}) with $T(E) = 1$.
The electron densities are calculated from the density of state
functions given by two times Eq. (\ref{eq.DE_hat}) for the Mexican hat dispersion and
by $m^*/\pi \hbar^2$ for the parabolic dispersion.
Over the range of Fermi energies shown, the electron density of the
Mexican hat dispersion is approximately 6 times larger than that of the
parabolic dispersion. 
To gain further insight, consider the itegrals of the density of states 
for low energies near the band edges, $n = \int_0^E dE' D(E')$. 
For the parabolic dispersion, $n_P = \frac{m^*}{\pi \hbar^2}E$, and for
the Mexican hat dispersion, $n_{MH} = \frac{4m^*}{\pi \hbar^2} \sqrt{\ep_0 E}$.
The ratio is $n_{MH} / n_P = 4 \sqrt{\ep_0/E}$. 
One factor of 2 results from the two branches of the Mexican hat dispersion at
low energies ($E<\ep_0$) and a second factor of 2 results from integrating
$1/\sqrt{E}$.
In this case, $\ep_0 = 0.111$ eV, so that at $E=0.05$ eV, $\sqrt{\ep_0/E}$
gives a factor of 1.5 resulting in a total factor of 6 
in the ratio $n_{MH} / n_P$ which is consistent with the numerical
calculation at finite temperature shown in Fig. \ref{fig:MexicanHat}(c).
There are two important points to take away from this plot. 
At the same electron density, the Fermi level of the Mexican hat
dispersion is much lower than that of the parabolic dispersion.
At the same electron density, the Seebeck coefficient of the Mexican
dispersion is much larger than the Seebeck coefficient of the 
parabolic dispersion.

Figure \ref{fig:MexicanHat}(d) compares the ballistic power factors 
calculated from the Mexican hat dispersion shown in Fig. \ref{fig:MexicanHat}(a)
and the parabolic dispersion, again with $m^* = m_0$ for both dispersions. 
The temperature is $T = 300$ K.
The ballistic power factor is calculated from Eqs. (\ref{eq:sigma}), 
(\ref{eq:S}), (\ref{eq:Ij}), and (\ref{eq:TE}) with $T(E) = 1$.
Eqs. (\ref{eq:DOM_2D}) and (\ref{eq.ME_hat}) 
for the density of modes are used
in Eq. (\ref{eq:TE}).
The peak power factor of the Mexican hat 
dispersion occurs when $E_F = -30.0$ meV, i.e. 30 meV below the 
conduction band edge.
This is identical to the analytical result obtained by
approximating the density of modes as an ideal step function.
The peak power factor of the parabolic dispersion
occurs when $E_F = -7.5$ meV.
At the peak power factors, 
the value of $I_{1}$ 
of the Mexican hat dispersion is 3.5 times larger than $I_1$
of the parabolic dispersion,
and $I_0$ of the Mexican hat dispersion 
is 3.2 times larger than $I_0$ of the parabolic dispersion.
The reason for the larger increase in $I_1$ compared
to $I_0$ is that, at the maximum power factor, 
the Fermi level of the Mexican hat dispersion
is further below the band edge.
Thus, the factor $(E - E_F)$ in the integrand of $I_1$
increases, and the average energy current referenced to the Fermi energy
given by $I_1$ increases more than the average particle current given
by $I_0$. 
%
%
Since the ratio $I_1 / I_0$ gives the Seebeck coefficient, this translates into
an increase of the Seebeck coefficient at the peak power factor.
At the peak power factors, the Seebeck coefficient 
of the Mexican hat dispersion is enhanced by 
10$\%$ compared to the parabolic dispersion.
We consistently observe a
larger increase in $I_1$ compared to that of $I_0$ 
at the peak power factor when comparing 
monolayer structures with Mexican hat dispersions to
bulk structures with parabolic dispersions.
For the III-VI materials, at their peak power factors,
GaSe shows a maximum increase of the Seebeck
coefficient of 1.4 between a monolayer with a Mexican hat dispersion
and bulk with a parabolic dispersion.
The power factor is proportional to $I_1^2/I_0 \propto S  I_1$.
Since the increase in $S$ at the peak power factor lies between 1 and 1.4,
the large increase in the maximum power factor results from the
large increase in $I_1$. 
Since the increase in $I_1$ is within a factor of 1 to 1.4 times the 
increase in $I_0$, one can also view the increase in the power
factor as resulting from an increase in $I_0$ which is simply the 
particle current or conductivity.
The increase of both of these quantities, $I_1$ or $I_0$,
results from the increase in the density of modes near the band edge
available to carry the current.
Over the range of integration of several $k_BT$ of the band edge,
the density of modes of the Mexican hat dispersion
is significantly larger than the density of modes 
of the parabolic dispersion as shown in Fig. \ref{fig:MexicanHat}(c).
%
%

From the Landauer-B{\"u}ttiker perspective of Eq. (\ref{eq:TE}),
the increased conductivity results from the increased number of 
modes.
From a more traditional perspective, the increased conductivity
results from an increased density of states resulting in an
increased charge density $n$.
At their peak power factors, the charge density of the Mexican hat dispersion
is $5.05 \times 10^{12}$ cm$^{-2}$, and 
the charge density of the parabolic dispersion is $1.57 \times 10^{12}$ cm$^{-2}$.
The charge density of the Mexican hat dispersion is 3.2 times larger
than the charge density of the parabolic dispersion
even though the Fermi level for the Mexican hat dispersion
is 22.5 meV less than the Fermi level of the parabolic dispersion. 
Since the peak power factor always occurs when $E_F$ is below
the band edge, the charge density resulting from the Mexican
hat dispersion will always be significantly larger than that of the
parabolic dispersion.
This, in general, will result in a higher conductivity.

When the height of the Mexican hat $\ep_0$ is reduced
by a factor of 4 ($k_0$ is reduced by a factor of 2), 
the peak power factor {\em decreases} by a factor of 2.5, 
the Fermi level at the peak power factor 
{\em increases} from -30 meV to -20.1 meV, 
and the corresponding electron density {\em decreases} by a factor 2.3.
When $\epsilon_{0}$ is varied with respect to the thermal energy at 300K 
using the following values, 5k$_{B}$T, 2k$_{B}$T,
k$_{B}$T and 0.5k$_{B}$T
the ratios of the Mexican hat power factors
with respect to the parabolic band power factors are 3.9, 2.2, 1.5
and 1.1, respectively.
The above analytical discussion illustrates the basic concepts
and trends, and it motivates the following numerical
investigation of various van der Waals materials exhibiting either Mexican hat
or Rashba dispersions.

\section{Computational Methods}
\label{sec:comp_methods}
Ab-initio calculations
of the bulk and few-layer structures (one to four layers) of GaS, GaSe, 
InS, InSe, Bi$_{2}$Se$_{3}$, Bi(111) surface, and bilayer graphene are carried out 
using density functional theory (DFT)
with a projector augmented wave method \cite{PAW} and the 
Perdew-Burke-Ernzerhof (PBE) type generalized gradient 
approximation \cite{perdew:1996:PBE,ernzerhof:1999:PBE_test} 
as implemented in the Vienna ab-initio Simulation Package (VASP). 
\cite{VASP1,VASP2}
The vdW interactions in 
 GaS, GaSe, InS, InSe and Bi$_{2}$Se$_{3}$
are accounted for using a 
semi-empirical correction to the Kohn-Sham energies when optimizing the bulk
structures of each material.\cite{Grimme_DFT_D2}
For the GaX, InX (X = S,Se), Bi(111) monolayer,
 and Bi$_{2}$Se$_{3}$ structures, a Monkhorst-Pack scheme is used for the integration of the Brillouin zone
with a k-mesh of 12 x 12 x 6 for the bulk structures and 12 x 12 x 1 for the
thin-films.
The energy cutoff of the plane wave basis is 300 eV.
The electronic bandstructure calculations
include spin-orbit coupling (SOC) for the GaX, InX, Bi(111) and Bi$_{2}$Se$_{3}$ compounds.
To verify the results of the PBE band structure calculations of the GaX and InX compounds,
the electronic structures of one to four monolayers of GaS and InSe are calculated
using the Heyd-Scuseria-Ernzerhof (HSE) functional.\cite{HSE_VASP}
The HSE calculations incorporate 25$\%$ short-range Hartree-Fock exchange.
The screening parameter $\mu$ is set to 0.2 \AA$^{-1}$.
For the calculations on bilayer graphene, a 32 $\times$ 32 $\times$ 1 k-point 
grid is used for the integration over the Brillouin zone.
The energy cutoff of the plane wave basis is 400 eV.
15\AA~of vacuum spacing was added to the slab geometries of all few-layer
structures.

The ab-initio calculations of the electronic structure
are used as input into a Landauer formalism for calculating the
thermoelectric parameters.
The two quantities requred are the density of states and the density of modes.
The density of states is directly provided by VASP.
The density of modes calculations are performed by integrating over the first
Brillouin zone using a converged k-point grid, $51 \times 51 \times 10$ k-points
for the bulk structures and $51 \times 51 \times 1$ k-points
for the III-VI, Bi$_{2}$Se$_{3}$ and Bi(111) thin film structures.
A $101 \times 101 \times 1$ grid of k-points is required
for the density of mode calculations on bilayer graphene.
The details of the formalism are provided in several prior 
studies.\cite{Lundstrom_Jesse_Bi2Te3,Klimeck_DOM_thermoelectric_JCE,Darshana_MX2_Thermo}
The temperature dependent carrier concentrations for each material
and thickness are calculated from the density-of-states obtained
from the ab-initio simulations.
To obtain a converged density-of-states a minimum k-point grid
of  72$\times$72$\times$36  (72$\times$72$\times$1)
is required for the bulk (monolayer and few-layer) III-VI and 
Bi$_{2}$Se$_{3}$ structures.
For the density-of-states calculations on bilayer graphene
and monolayer Bi(111) a 36$\times$36$\times$1 grid
of k-points is used.

The calculation of the conductivity, the power factor, 
and $ZT$ requires values for the electron and hole
mean free paths and the lattice thermal conductivity.
Electron and hole scattering are included
using a constant mean free path, $\lambda_{0}$ determined by
fitting to experimental data.
For GaS, GaSe, InS and InSe, $\lambda_{0}$ = 25 nm 
gives the best agreement with experimental data. \cite{GaSe_sigma_PSSb,
GaS_sigma_JAP, InSe_sigma_TSF, InS_sigma_JPhysChem}
The room temperature bulk n-type electrical conductivity of GaS, 
GaSe, InS and InSe at room 
temperature was reported to be 0.5 $\Omega^{-1}$m$^{-1}$, 
0.4 $\Omega^{-1}$m$^{-1}$, 0.052 $\Omega^{-1}$m$^{-1}$ and 0.066 $\Omega^{-1}$m$^{-1}$ 
respectively at a carrier concentration of 10$^{16}$ cm$^{-3}$.
Using $\lambda_{0}$ = 25 nm for bulk GaS, GaSe and InSe
we obtain an electrical conductivity
of 0.58 $\Omega^{-1}$m$^{-1}$, 0.42 $\Omega^{-1}$m$^{-1}$,
0.058 $\Omega^{-1}$m$^{-1}$ 
and 0.071 $\Omega^{-1}$m$^{-1}$, respectively at the same
carrier concentration.
For the Bi(111) monolayer surface, the relaxation time for scattering
in bulk Bi is reported to be 0.148 ps at 300K.\cite{Bi_1L_thermo_JPC}
Using the group velocity of the conduction and valence bands 
($\sim 6.7 \times 10^{4}$ m/sec for electrons and holes)
from our ab-initio simulations, an electron and hole mean free
path of 10 nm is used to determine the thermoelectric parameters
of the Bi(111) monolayer.
Prior theoretical studies of scattering in thin films of Bi$_{2}$Se$_{3}$
ranging from  2 QLs to 4 QLs give a scattering time on 
the order of 10 fs.\cite{GYin_TI_DRC, GYin_TI_APL14,Udo_Bi2Se3}
Using a scattering time of $\tau$ = 10 fs and electron
and hole group velocities from the ab-initio simulations
of 3 $\times$10$^{5}$ m/s and
2.4$\times$10$^{5}$ m/s, respectively,
electron and hole mean free paths of $\lambda_{e}$=3 nm and $\lambda_{p}$=2.4 nm
are used to extract the thermoelectric parameters for bulk and
thin film Bi$_{2}$Se$_{3}$.
For bilayer graphene, $\lambda_{0}$ = 88 nm 
gives the best agreement with experimental data on 
conductivity at room temperature.\cite{bilayergraphene_thermopower_PRL11}

Values for the 
lattice thermal conductivity are also taken from
available experimental data.
The thermal conductivity in defect-free thin films is limited
by boundary scattering and can be up to an order of magnitude
lower than the bulk thermal conductivity.\cite{Lundstrom_Si_thermal}
As the thickness of the film increases, $\kappa_{l}$ approaches
the Umklapp limited thermal conductivity of the bulk structure.
Hence, the values of $\kappa_{l}$ obtained from experimental studies
of bulk materials for this study are
an upper bound approximation of $\kappa_{l}$ in 
the thin film structures.
The experimental value of 10 Wm$^{-1}$K$^{-1}$ reported for the
in-plane lattice thermal conductivity $\kappa_{l}$ of
bulk GaS at room temperature 
is used for the gallium chalcogenides.\cite{GaS_kappa_exp}
The experimental, bulk, in-plane, lattice
thermal conductivities of 
7.1 Wm$^{-1}$K$^{-1}$ and 12.0 Wm$^{-1}$K$^{-1}$
measured at room temperature
are used for InS and InSe, respectively.
\cite{Spitzer_kappaL_JPhysChem}
For monolayer Bi(111), 
the calculated $\kappa_{l}$ 
from molecular dynamics \cite{Bi_1L_thermo_JPC} at 300K 
is 3.9 Wm$^{-1}$K$^{-1}$. 
For Bi$_{2}$Se$_{3}$, 
the measured bulk $\kappa_{l}$ value 
at 300K is 2 Wm$^{-1}$K$^{-1}$.\cite{goldsmid2009thermo, Bi2Se3_Cava_PRB}
A value of 2000  Wm$^{-1}$K$^{-1}$
is used for the room temperature 
in-plane lattice thermal conductivity
of bilayer graphene.
This is consistent with a number of 
experimental measurements and theoretical
predictions on the lattice thermal conductivity of bilayer graphene.
\cite{balandin2011thermal, KWKim_graphene_kappa}
When evaluating $ZT$ in Eq. (\ref{eq:ZT}) for the 2D, thin film structures, 
the bulk lattice thermal conductivity is multiplied by the film thickness. 
When tabulating values of the electrical conductivity 
and the power factor of the 2D films, 
the calculated conductivity from Eq. (\ref{eq:sigma}) is divided
by the film thickness.

Much of the experimental data from which the 
values for $\lambda_0$ and $\kappa_l$ have been determined 
are from bulk studies,
and clearly these values might change as the
materials are thinned down to a few monolayers.
However, there are presently no experimental values available
for few-layer III-VI and Bi$_2$Se$_3$ materials.
Our primary objective is to obtain a qualitative understanding
of the effect of the bandstructure in these materials
on their thermoelectric properties.
To do so, we 
use the above values for $\lambda_0$
and $\kappa_l$ to calculate $ZT$
for each material as a function of thickness.
We tabulate these values and provide the corresponding values
for the electron or hole density, Seebeck coefficient, and conductivity
at maximum $ZT$.
It is clear from Eqs. (\ref{eq:S}) and (\ref{eq:Ij})
that the Seebeck coefficient is relatively insensitive to the 
value of the mean free path.
Therefore, when more accurate values for the conductivity or
$\kappa_l$ become available,
new values for $ZT$ can be estimated from
Eq. (\ref{eq:ZT}) using
the given Seebeck coefficient and replacing the 
electrical and/or thermal conductivity.

\section{Numerical Results}
\label{sec:num_results}
\subsection{III-VI Compounds GaX and InX (X = S, Se)}
\label{sec:GaX_InX}
The lattice parameters of the optimized bulk GaX and InX compounds
are summarized in Table \ref{tab:mat_params}.
For the GaX and InX compounds the lattice parameters and bulk bandgaps obtained
are consistent with prior experimental \cite{Kuhn_GaS_experiment, Kuhn_GaSe_experiment}
and theoretical studies \cite{zolyomi_GaX, zolyomi_InX,GaSe_GaS_1L_PCCP} of the bulk crystal structure
and electronic band structures.
\begin{table*}
\begin{tabular}{p{2cm} p{1.6cm} p{1.6cm} p{1.6cm} p{1.6cm} p{1.6cm} p{1.6cm} p{1.6cm} p{1.6cm} p{1.6cm}}
\hline\hline
& $a_{0}$(\AA) &  $c_{0}$(\AA)  & $d$ (\AA) & $d_{vdW}$ (\AA) & $a_{0}^{expt}$(\AA) & $c_{0}^{expt}$(\AA) & $d^{expt}$ (\AA)  & $E_{g}$(eV) & $E_{g}^{expt}$(eV) \\[0.5ex]  \hline
GaS & 3.630 & 15.701 &  4.666 & 3.184 & 3.587  & 15.492  & 4.599 & 1.667 &  -\\ [0.5ex]
GaSe & 3.755 & 15.898 &  4.870 & 3.079 &  3.752  & 15.950  & 4.941 & 0.870 & 2.20 \\ [0.5ex]
InS  & 3.818 & 15.942 &  5.193 &  2.780 & \dots  & \dots  & \dots & 0.946 & - \\ [0.5ex]
InSe & 4.028 & 16.907 &  5.412 & 3.040& 4.000 & 16.640  & 5.557 & 0.48 & 1.20 \\ [0.5ex]
\hline 
Bi$_{2}$Se$_{3}$ & 4.140 & 28.732 &  7.412 & 3.320 &  4.143 & 28.636  & \dots & 0.296 & 0.300 \\ [0.5ex]
\hline
BLG & 2.459 & - &  3.349 & 3.349 & 2.460 & -& 3.400 & - & - \\ [0.5ex]
\hline
Bi(111) & 4.34 & - &  3.049 & - &4.54 & -& - & 0.584 & - \\ [0.5ex]
\hline
\end{tabular}
\caption{Calculated properties of bulk Mexican-hat materials GaS, GaSe, InS, InSe, Bi$_{2}$Se$_{3}$,
 bilayer graphene (BLG), and Bi(111) : lattice constant $a_{0}$, c-axis lattice constant $c_{0}$, thickness of individual
layer $d$ and bandgap $E_{g}$(eV).  The calculated thickness, $d$, is the atom-center to atom-center distance
between the top and bottom chalcogen atoms of a single layer in  GaS, GaSe, InS, InSe, Bi$_{2}$Se$_{3}$ and
atom center to atom center distance of the top and bottom carbon atoms in bilayer graphene.  The thickness, $d$
in monolayer Bi is the height of the buckling distance between the two Bi atoms.
Experimental values when available
\cite{Kuhn_GaS_experiment, Kuhn_GaSe_experiment, Blasi_InSe_experiment,
Ajayan_InSe,nakajima_Bi2Se3_structure_expt, Ohta_bilayer_G} 
are included for comparison.}
\label{tab:mat_params} 
\end{table*}

\begin{figure}[!h]
\includegraphics[width=5in]{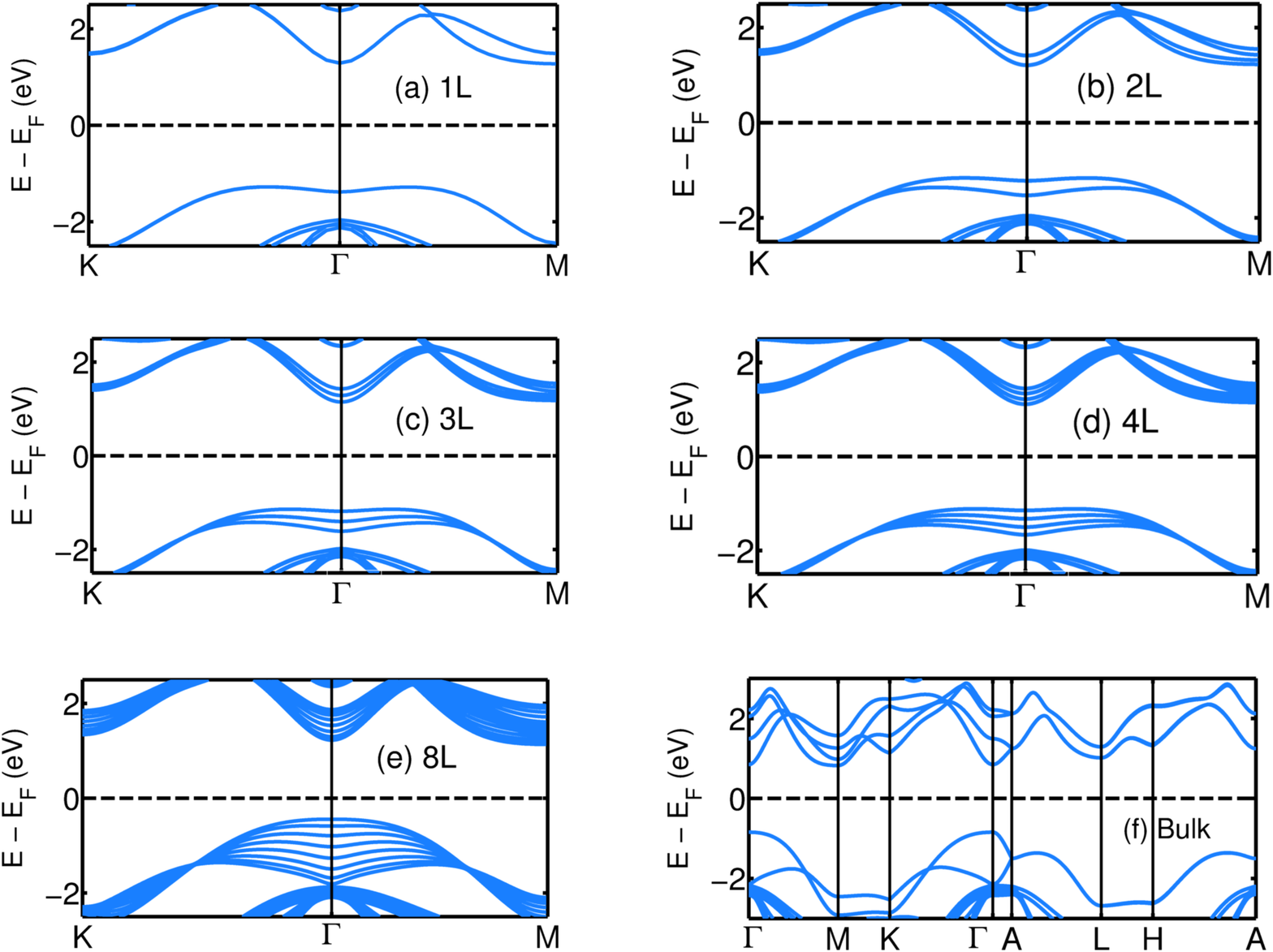}
\caption{(Color online) PBE SOC band structure of GaS: (a) 1L,
(b) 2L, (c) 3L and (d) 4L, (e) 8L and (f) bulk GaS. 
}
\label{fig:Ek_GaS}
\end{figure}

In this study, the default stacking is the  $\beta$ phase illustrated
in Fig. \ref{fig:MexicanHat}a.
The $\beta$ phase 
is isostructural to the AA' stacking order in the 2H polytypes of the
molybdenum and tungsten dichalcogenides.\cite{Franchini_stacking_PRB14}
The bandgap of the one to four monolayer structures is indirect
for GaS, GaSe, InS and InSe.
Figure \ref{fig:Ek_GaS} illustrates the PBE band structure for one-layer (1L)
through four-layers (4L), eight-layer (8L) and bulk GaS.
The PBE SOC band gaps and energy transitions 
for each of the III-VI materials and film thicknesses are
are listed in Table \ref{tab:GaX_InX_Egap}. 
For GaS, the HSE SOC values are also listed.
The effective masses 
extracted from the PBE SOC electronic bandstructure 
are listed in Table \ref{tab:eff_mass}.
\begin{table*}
\begin{tabular} {c | c |  c c c c}

\hline\hline
Structure & Transition &  GaS  & GaSe & InS & InSe \\[0.5ex]  \hline
1L & E$_{v}$ to $\Gamma_{c}$ & 2.563 (3.707)& \textbf{2.145} & \textbf{2.104} & \textbf{1.618}  \\ [0.5ex] 
   & E$_{v}$ to $K_{c}$ & 2.769 (3.502) & 2.598 & 2.684 & 2.551   \\ [0.5ex] 
   &E$_{v}$  to $M_{c}$ & \textbf{2.549 (3.422)} & 2.283 & 2.520 & 2.246   \\ [0.5ex] \hline

2L &E$_{v}$  to $\Gamma_{c}$ & \textbf{2.369 (3.156) }& \textbf{1.894}  & \textbf{1.888} & \textbf{1.332}  \\ [0.5ex] 
   &E$_{v}$  to $K_{c}$ & 2.606 (3.454) & 2.389 & 2.567 & 2.340   \\ [0.5ex] 
   &E$_{v}$ to $M_{c}$ & 2.389 (3.406) & 2.065 & 2.353 & 2.025 \\ [0.5ex] \hline

3L &E$_{v}$ to $\Gamma_{c}$ & \textbf{2.288 (3.089)} & \textbf{1.782} & \textbf{1.789} & \textbf{1.152}  \\ [0.5ex] 
   &E$_{v}$ to $K_{c}$ & 2.543 (3.408) & 2.302 & 2.496 & 2.201   \\ [0.5ex] 
   &E$_{v}$ to $M_{c}$ & 2.321 (3.352) & 1.967 & 2.273 & 1.867  \\ [0.5ex] \hline

4L &E$_{v}$ to $\Gamma_{c}$ & \textbf{2.228 (3.011)} & \textbf{1.689} & \textbf{1.749} & \textbf{1.086} \\ [0.5ex] 
   &E$_{v}$ to $K_{c}$ & 2.496 (3.392) & 2.224 &  2.471 & 2.085   \\ [0.5ex] 
   &E$_{v}$ to $M_{c}$ & 2.267 (3.321) & 1.879 &2.242 & 1.785  \\ [0.5ex] \hline

   Bulk &$\Gamma_{v}$ to $\Gamma_{c}$ & 1.691 (2.705) & \textbf{0.869} & \textbf{0.949} & \textbf{0.399} \\ [0.5ex] 
   &$\Gamma_{v}$ to $K_{c}$ & 1.983 (2.582) & 1.435 &  1.734 & 1.584   \\ [0.5ex] 
   &$\Gamma_{v}$ to $M_{c}$ & \textbf{1.667} (\textbf{2.391})  & 0.964 & 1.400 & 1.120  \\ [0.5ex] 
\hline
\end{tabular}
\caption{ PBE SOC calculations of the bandgap energies and energy transitions
between the valence band edge of the Mexican hat band (E$_{v}$) 
and the conduction ($c$) band valleys for 1L to 4L GaS, GaSe, InS and InSe. 
 The bandgap at each dimension is highlighted in bold text.  The HSE-SOC
energy transitions for GaS are in parentheses.  }
\label{tab:GaX_InX_Egap}
\end{table*}
\begin{table*}
\begin{tabular} {c | p{1.5cm} p{1.5cm} p{1.5cm} p{1.5cm}| p{2.7cm} p{1.5cm} p{1.5cm} p{1.5cm}}
\hline\hline
Structure &  GaS  & GaSe & InS & InSe &  GaS  & GaSe & InS & InSe  \\[0.5ex]  \hline
& \multicolumn{4}{c|}{Hole Effective Mass (m$_{0}$)} & \multicolumn{4}{c}{Electron Effective Mass (m$_{0}$)} \\[0.5ex] \hline
1L & 0.409 & 0.544 & 0.602 & 0.912 & 0.067 (0.698) & 0.053 & 0.080 & 0.060\\ [0.5ex]
    
2L & 0.600 & 0.906  & 0.930 & 1.874  & 0.065 (0.699) & 0.051 & 0.075 & 0.055  \\ [0.5ex]
    
3L & 0.746 & 1.439 & 1.329  & 6.260 & 0.064 (0.711) & 0.050 & 0.074 & 0.053 \\ [0.5ex]
    
4L & 0.926 & 2.857  & 1.550 & 3.611 & 0.064 (0.716) & 0.049 & 0.073 & 0.055 \\ [0.5ex]
    
\hline
\end{tabular}
\caption{Ab-initio calculations of the hole and electron effective masses
at the $\Gamma$ valley of the valence band and conduction band respectively
for each structure in units of the free electron mass (m$_{0}$).  The conduction band
effective masses at M$_{c}$ are included in parentheses for one to four layers of GaS.  }
\label{tab:eff_mass} 
\end{table*}

The conduction bands of GaSe, InS, and InSe are at $\Gamma$ for
all layer thicknesses, from monolayer to bulk.
The conduction band of monolayer GaS is at M. 
This result is consistent with that of Z\'{o}lyomi et al.\cite{zolyomi_GaX}. 
However, for all thicknesses
greater than a monolayer, the conduction band of GaS is at $\Gamma$.
Results from the PBE functional give GaS conduction valley separations between
M and $\Gamma$ that are on
the order of $k_BT$ at room temperature, and this leads to qualitatively
incorrect results in the calculation of the electronic and thermoelectric 
parameters.
For the three other III-VI compounds, the minimum PBE-SOC spacing
between the conduction $\Gamma$ and M valleys is 138 meV in 
monolayer GaSe. 
For InS and InSe, the minimum conduction $\Gamma$-M valley
separations also occur for a monolayer, and they are
416 eV and 628 eV, respectively.
For monolayer GaS, the HSE-SOC conduction M valley lies 80 meV 
below the K valley and 285 meV below the $\Gamma$ valley.
At two to four layer thicknesses, the order is reversed, 
the conduction band edge is at $\Gamma$, 
and the energy differences between the valleys increase.
For the electronic and thermoelectric properties, only energies within
a few $k_BT$ of the band edges are important.
Therefore, the density of modes of n-type GaS is calculated from the HSE-SOC
bandstructure.
For p-type GaS and all other materials, the densities of modes
are calculated from the PBE-SOC bandstructure.

The orbital composition of the monolayer GaS conduction
$\Gamma$ valley contains 63\% Ga $s$ orbitals
and 21\% S $p_z$ orbitals. 
The orbital compositions of the other III-VI conduction $\Gamma$
valleys are similar.
As the film thickness increases from a monolayer to a bilayer, 
the conduction $\Gamma$ valleys in each layer couple and split by 203 meV
as shown in Fig. \ref{fig:Ek_GaS}b.
Thus, as the film thickness increases,
the number of low-energy $\Gamma$ states near the conduction band-edge remains the same,
or, saying it another way,
the number of low-energy $\Gamma$ states per unit thickness decreases
by a factor of two
as the the number of layers increases from a monolayer to a bilayer.
This affects the electronic and thermoelectric properties.

The Mexican hat feature of the valence band is present in all of the 
1L - 4L GaX and InX structures, 
and it is most pronounced for the monolayer structure
shown in Fig. \ref{fig:Ek_GaS}a.
For monolayer GaS, the highest valence band at $\Gamma$ is 
composed of 79\% sulfur $p_z$ orbitals ($p_z^S$).
The lower 4 valence bands at $\Gamma$ 
are composed entirely of sulfur $p_x$ and $p_y$ 
orbitals ($p_{xy}^S$).
When multiple layers are brought together, the 
$p_z^S$ valence band at $\Gamma$ strongly couples and splits
with a splitting of 307 meV in the bilayer.
For the 8-layer structure in Fig. \ref{fig:Ek_GaS}e, the manifold of
8 $p_z^S$ bands touches the manifold
of $p_{xy}^S$ bands, and the bandstructure
is bulklike with discrete $k_z$ momenta.
In the bulk shown in Fig. \ref{fig:Ek_GaS}f, the discrete energies become a continuous dispersion
from $\Gamma$ to $A$.
At 8 layer thickness, the large splitting of the
$p_z^S$ valence band removes the Mexican hat feature,
and the valence band edge is parabolic as in the bulk.
The nature and orbital composition of the bands of the 4
III-VI compounds are qualitatively the same.

\begin{table}[!h]
\begin{tabular} {l | c |  c}
\hline\hline
Material & $\ep_0$ (meV) &  $k_0$ (nm$^{-1}$)\\  
(Theory/Stacking Order) &  1L, 2L, 3L, 4L & 1L, 2L, 3L, 4L   \\
\hline
GaS      & 111.2, 59.6, 43.8, 33.0 & 3.68, 2.73,  2.52, 2.32\\
GaS (no-SOC)     & 108.3, 60.9, 45.1, 34.1 & 3.16, 2.63, 2.32, 2.12\\
GaS (HSE) & 97.9, 50.3, 40.9, 31.6 & 2.81, 2.39, 2.08, 1.75\\
GaS (AA)  & 111.2, 71.5, 57.1, 47.4 & 3.68, 2.93, 2.73, 2.49\\
GaSe     & 58.7, 29.3, 18.1, 10.3  & 2.64, 2.34, 1.66, 1.56\\
GaSe ($\epsilon$) & 58.7, 41.2, 23.7 , 5.1   & 2.64, 1.76, 1.17 , 1.01\\
InS      & 100.6, 44.7, 25.8, 20.4  & 4.03, 3.07, 2.69, 2.39\\
InSe     & 34.9, 11.9, 5.1, 6.1    & 2.55, 1.73, 1.27, 1.36\\
InSe (HSE) & 38.2, 15.2, 8.6, 9.2 & 2.72, 2.20, 1.97, 2.04\\
Bi$_2$Se$_3$ & 314.7, 62.3, 12.4, 10.4 & 3.86, 1.23, 1.05, 0.88\\
Bi$_2$Se$_3$ (no-SOC) & 350.5, 74.6, 22.8, 20.1 & 4.19, 1.47, 1.07, 1.02\\

\hline
\end{tabular}
\caption{
Values of $\ep_0$ and $k_0$ are listed in order of 
thicknesses: 1L, 2L, 3L, and 4L. 
The default level of theory is PBE with 
spin-orbit coupling, and the default stacking is AA'.
Only deviations from the defaults are noted.
}
\label{tab:e0_k0_III_VIs}
\end{table}
In the few-layer structures, 
the Mexican hat feature of the valence band can be characterized by the
height, $\epsilon_{0}$, at $\Gamma$ and the radius of the band-edge ring, $k_0$,
as illustrated in Figure \ref{fig:MexicanHat}(b).
The actual ring has a small anisotropy that has been previously characterized
and discussed in detail \cite{zolyomi_GaX,zolyomi_InX,SGLouie_GaSe_arxiv}.
For all four III-VI compounds of monolayer and few-layer thicknesses,
the valence band maxima (VBM)
of the inverted Mexican hat lies along $\Gamma -K$, 
and it is slightly
higher in energy compared to the band extremum along $\Gamma - M$.
In monolayer GaS, the valence band maxima along $\Gamma - K$ is 
4.7 meV above the band extremum along $\Gamma - M$.
In GaS, as the film thickness increases from one layer to four layers
the energy difference between the two extrema 
decreases from 4.7 meV to 0.41 meV.
The maximum energy difference of 6.6 meV
between the band extrema of the Mexican hat occurs in a
monolayer of InS.
In all four III-VI compounds the energy difference between the band extrema
is maximum for the monolayer structure and decreases below 0.5 meV in all of 
the materials for the four-layer structure.
The tabulated values of $k_0$ in
Table \ref{tab:e0_k0_III_VIs} 
give the distance from $\Gamma$ to the VBM in the $\Gamma - K$
direction.
Results calculated from PBE and HSE functionals are given, 
and results with and without spin-orbit coupling are listed. 
The effects of AA' versus AA stacking order of GaS and
AA' versus $\epsilon$ stacking order of GaSe \cite{GaSe_topological_Udo,
Zheng_GaSe_topological_JCP} are also compared.
Table \ref{tab:e0_k0_III_VIs} shows that the valence band Mexican hat feature is robust.
It is little affected by the choice of functional, the omission or inclusion of
spin-orbit coupling, or the stacking order.
A recent study of GaSe at the G$_0$W$_0$ level found that the Mexican hat
feature is also robust against many-electron self-energy 
effects.\cite{SGLouie_GaSe_arxiv}
For all materials, the values of $\epsilon_{0}$ and $k_0$ are 
largest for monolayers 
and decrease as the film thicknesses increase.
This suggests that the height of the step function density of modes
will also be maximum for the monolayer structures. 
\begin{figure}[t]
\includegraphics[width=5in]{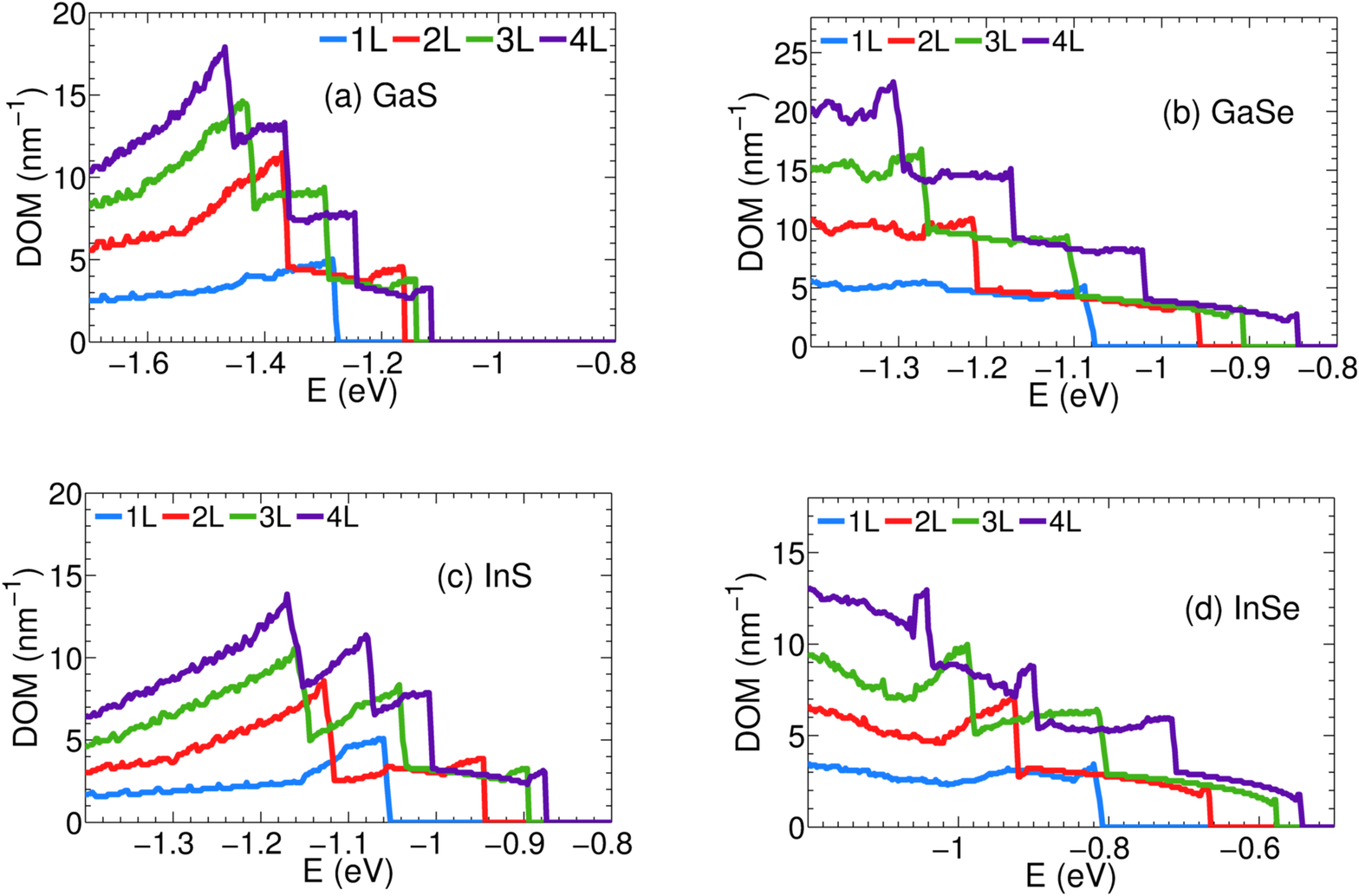}
 \caption{(Color online) Distribution of valence band modes per unit width versus energy
 for (a) GaS, (b) GaSe, (c) InS and (d) InSe for 1L (blue), 2L (red), 
 3L (green) and 4L (purple) structures.  The midgap energy is set to E=0.
 }
\label{fig:GaX_DOM}
\end{figure}
Figure \ref{fig:GaX_DOM} illustrates the valence band density of modes for 1L, 2L, 3L and 4L
GaS, GaSe, InS and InSe.
The valence band density of modes is a step function for the few-layer
structures, and the height of the step function at the valence band edge is 
reasonably approximated by Eq. (\ref{eq.ME_bandedge}).
The height of the numerically calculated 
density of modes step function for monolayer
GaS, GaSe, InS and InSe is 4.8 nm$^{-1}$, 5.2 nm$^{-1}$,
5.1 nm$^{-1}$ and 3.4 nm$^{-1}$ respectively.
Using the values for $k_0$ and Eq. (\ref{eq.ME_bandedge})
and accounting for spin degeneracy,
the height of the step function for monolayer 
GaS, GaSe, InS and InSe is 4.1 nm$^{-1}$, 3.4 nm$^{-1}$,
5.1 nm$^{-1}$ and 3.2 nm$^{-1}$.
The height of the numerically calculated density of modes in GaS
decreases by $\sim30\%$ when the film thickness increases from one to 
four monolayers, and the value of $k_0$ decreases by $\sim38\%$.
The height of the step function using Eq. (\ref{eq.ME_bandedge}) and $k_0$
is either underestimated or equivalent to the numerical density of modes.
For all four materials GaS, GaSe, InS and InSe, decreasing the film
thickness increases $k_0$ and the height of the 
step-function of the band-edge density of modes.
A larger band-edge density of modes gives a
larger power factor and ZT compared to that of the bulk.
\mbox{ }
\begin{figure}[t]
\includegraphics[width=6.5in]{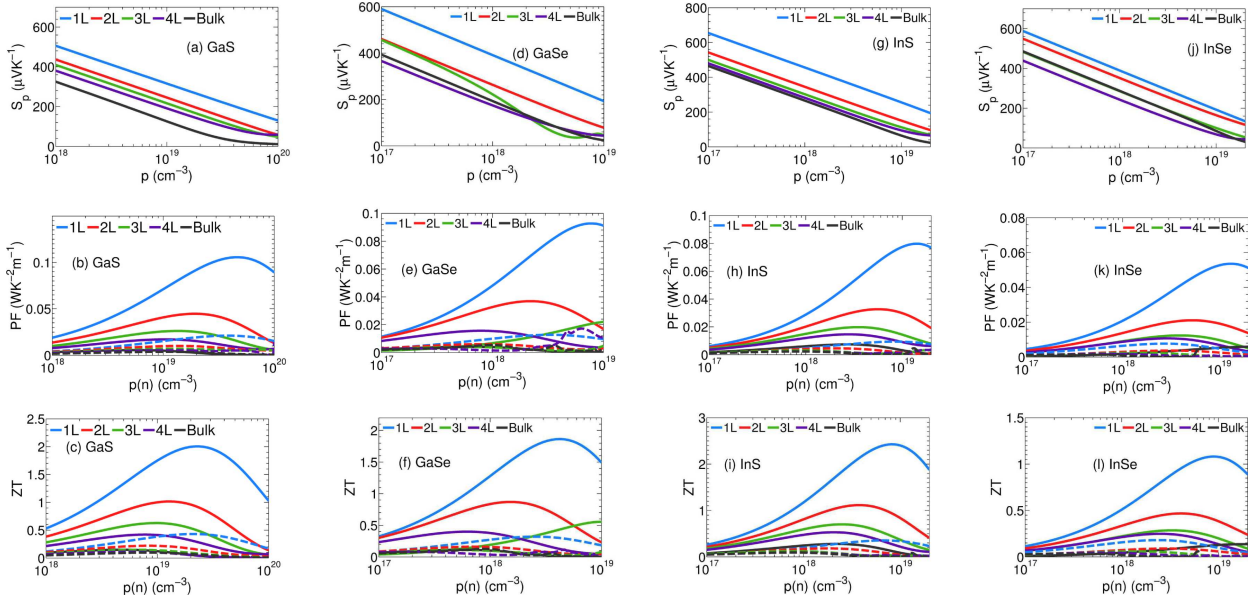}
\caption{(Color online) Seebeck coefficient, power factor and
thermoelectric figure-of-merit, ZT, of p-type (solid line)
and n-type (broken line) 1L (blue), 2L (red), 3L (green), 4L (purple) and bulk (black)
(a)-(c) GaS, (d)-(f) GaSe, (g)-(i) InS and (j)-(l) InSe at room temperature.}
\label{fig:III_VI_thermo}
\end{figure}
The p-type Seebeck coefficients, the p-type and n-type 
power factors, and the thermoelectric figures-of-merit (ZT) 
as functions of carrier concentration at room temperature
for GaS, GaSe, InS and InSe
are shown in Figure \ref{fig:III_VI_thermo}.
The thermoelectric parameters at $T=300$ K of bulk and one to four monolayers 
of GaS, GaSe, InS and InSe are summarized in Tables 
\ref{tab:GaS_thermo} - \ref{tab:InSe_thermo}.
For each material the peak p-type 
ZT occurs at a monolayer thickness.
The largest room temperature p-type ZT occurs in monolayer InS. 
At room temperature, 
the peak p-type (n-type) ZT values in 1L, 2L, 3L and 4L GaS
occur when the Fermi level is 42 meV, 38 meV, 34 meV and
30 meV (22 meV, 17 meV, 11 meV, and 7 meV) above (below)
the valence (conduction) band edge,
and
the Fermi level positions in GaSe, InS and InSe 
change in qualitatively the same way.
The p-type hole concentrations of monolayer GaS, GaSe, InS and InSe
at the peak ZT are enhanced by factors of
9.7, 10.8, 7.2 and 5.5 compared to those of their respective bulk structures.
At the peak p-type room-temperature ZT, the Seebeck coefficients of 
monolayer GaS, GaSe, InS and InSe
are enhanced by factors of 1.3, 1.4, 1.3, and 1.3, respectively, 
compared to their bulk values.
However, the monolayer and bulk peak ZT values occur at carrier concentrations that differ by an
order of magnitude.
At a fixed carrier concentration, the monolayer Seebeck coefficients
are approximately 3.1 times larger than the bulk Seebeck coefficients.
The p-type power factor (PF) at the peak ZT for 1L GaS is enhanced by a factor of 17 
compared to that of bulk GaS.
The p-type ZT values 
of monolayer GaS, GaSe, InS and InSe 
are enhanced by 
factors of 14.3, 16.9, 8.7 and 7.7, respectively, 
compared to their bulk values.
At the peak p-type ZT, 
the contribution of $\kappa_{e}$ to $\kappa_{tot}$ is 
minimum for the bulk structure and
maximum for the
monolayer structure.
The contributions of $\kappa_{e}$ to $\kappa_{tot}$ 
in bulk and monolayer GaS are 5\% and 24\%, respectively. 
The increasing contribution of $\kappa_{e}$ to $\kappa_{tot}$ with
decreasing film thickness reduces the enhancement of ZT relative to that of
the power factor.

\begin{centering}
\vspace{0.2in}
\begin{table*}
\begin{tabular}{l c c c c | c c c c}
\hline\hline
Thickness & \multicolumn{1}{c}{p} & \multicolumn{1}{c}{$S_{p}$} &  \multicolumn{1}{c}{$ \sigma_{p}$}  &  \multicolumn{1}{c}{ZT$_{p}$} &\multicolumn{1}{c}{n} & \multicolumn{1}{c}{$\mid$|S$_{e}\mid$} &  \multicolumn{1}{c}{$\sigma_{e}$}  &  \multicolumn{1}{c}{ZT$_{e}$} \\[0.5ex]  
  & \multicolumn{1}{c}{($\times$10$^{19}$cm$^{-3}$)} & \multicolumn{1}{c}{($\mu VK^{-1}$)} &  \multicolumn{1}{c}{($\times 10^{6}\Omega m)^{-1}$}  & \multicolumn{1}{c}{ } & \multicolumn{1}{c}{($\times$10$^{19}$ cm$^{-3}$)} & \multicolumn{1}{c}{($\mu VK^{-1}$)} &  \multicolumn{1}{c}{($\times 10^{6}\Omega m)^{-1}$}   &  \multicolumn{1}{c}{ } \\[0.5ex] 
\hline
1L  &  3.19  & 251.6 & 1.41 &  2.01 & 1.02 & 237.0 &  .348 & .431 \\ [0.5ex] \hline

2L  & 1.51 & 222.9 & .776 &  1.02 & .621  & 219.6 &  .229 & .218 \\ [0.5ex] \hline

3L  & 1.13 & 213.2 & .530 &  .630 & .595 & 200.9 &  .206 & .147 \\ [0.5ex] \hline

4L  & .922 & 211.2 & .390 &  .421 & .545 & 191.9 & .195 & .111 \\ [0.5ex] \hline

Bulk  & .330 & 187.6 & .149 & .140 & .374 & 210.8 &  .116 & .095 \\  [0.5ex] \hline
\hline 
\end{tabular}
\caption{ 
GaS thermoelectric properties for bulk and one to four monolayers at 
$T=300$ K. 
Hole and electron carrier concentrations (p and n), Seebeck coefficients (S$_{p}$ and S$_{e}$), 
and electrical conductivties ($\sigma_{p}$ and $\sigma_{n}$) at
the peak p-type and n-type ZT.}
\label{tab:GaS_thermo}
\end{table*}
%
\end{centering}

The increases in the Seebeck coefficients, the charge densities, 
and the electrical conductivities with decreases in
the film thicknesses follow the 
increases in the magnitudes of $I_{0}$ and $I_{1}$
as discussed at the end of Sec. \ref{sec:analytical}.
For bulk p-type GaS,
the values of $I_0$ ($I_1$) at peak ZT are 0.94 (1.85), and
for monolayer GaS, they are 8.87 (23.4). 
They increase by factors of 9.4 (12.6) as the film thickness
decreases from bulk to monolayer.
In 4L GaS, the values of $I_0$ ($I_1$) are 2.45 (5.38), and
they increase by factors of 3.6 (5.4) as the thickness
is reduced from 4L to 1L.
For all four of the III-VI compounds, the increases in $I_{1}$ are 
larger than the increases in $I_0$ as the film thicknesses decrease.
As described in Sec. \ref{sec:analytical}, these increases
are driven by the transformation of the dispersion from parabolic 
to Mexican hat with an increasing radius of the band edge $k$-space
ring as the thickness is reduced from bulk to monolayer.

\begin{centering}
\vspace{0.2in}
\begin{table*}
\begin{tabular}{l c c c c | c c c c}
\hline\hline
Thickness & \multicolumn{1}{c}{p} & \multicolumn{1}{c}{$S_{p}$} &  \multicolumn{1}{c}{$ \sigma_{p}$}  &  \multicolumn{1}{c}{ZT$_{p}$} &\multicolumn{1}{c}{n} &\multicolumn{1}{c}{$\mid$|S$_{e}\mid$} &  \multicolumn{1}{c}{$\sigma_{e}$}  &  \multicolumn{1}{c}{ZT$_{e}$} \\[0.5ex]  
  & \multicolumn{1}{c}{($\times$10$^{18}$cm$^{-3}$)} & \multicolumn{1}{c}{($\mu VK^{-1}$)} &  \multicolumn{1}{c}{($\times 10^{6}\Omega m)^{-1}$}  & \multicolumn{1}{c}{ } & \multicolumn{1}{c}{($\times$10$^{18}$ cm$^{-3}$)} & \multicolumn{1}{c}{($\mu VK^{-1}$)} &  \multicolumn{1}{c}{($\times 10^{6}\Omega m)^{-1}$}   &  \multicolumn{1}{c}{ } \\[0.5ex] 
\hline
1L    & 5.81 & 256.1 & 1.28 &  1.86 & 2.71 & 202.9 &  .310 & .321 \\ [0.5ex] \hline

2L    & 2.70 & 225.3 & .711 &  .870 & 1.20 & 201.4 &  .152 & .162 \\ [0.5ex] \hline

3L    & 2.09 & 221.2 & .450 & .561 & .79 & 194.0 &  .103 & .110 \\ [0.5ex] \hline

4L    & 1.49 & 210.2 & .352 &  .391 & .69 & 186.4 &  .085 & .082 \\ [0.5ex] \hline

Bulk  & .541 & 180.9 & .121 & .112 & .29 & 127.9 &  .033 & .132 \\ [0.5ex] \hline

\hline 
\end{tabular}
\caption{ 
GaSe
thermoelectric properties for bulk and one to four monolayers at 
300K.
Hole and electron carrier concentrations (p and n), Seebeck coefficients (S$_{p}$ and S$_{e}$), 
and electrical conductivties ($\sigma_{p}$ and $\sigma_{n}$) at
the peak p-type and n-type ZT.
}
\label{tab:GaSe_thermo}
\end{table*}
%
\end{centering}

While the focus of the paper is on the effect of the 
Mexican hat dispersion that forms in the
valence band of these materials, the n-type thermoelectric figure of merit 
also increases as the film thickness is reduced to a few layers,
and it is also maximum at monolayer thickness.
The room temperature, monolayer, n-type thermoelectric figures of merit 
of GaS, GaSe, InS and InSe
are enhanced by factors of 4.5, 2.4, 3.8 and 5.3, 
respectively, compared to the those of the respective bulk structures.
The largest n-type ZT occurs in monolayer GaS.
In a GaS monolayer, the 3-fold degenerate M valleys form the
conduction band edge.
This large valley degeneracy gives GaS the largest n-type ZT
among the 4 III-VI compounds.
As the GaS film thickness increases from 
a monolayer to a bilayer, the conduction band edge moves to the 
non-degenerate $\Gamma$ valley so that the number of low-energy
states near the conduction band edge decreases.
With an added third and fourth layer, 
the M valleys move higher, 
and the $\Gamma$ valley continues to split 
so that the number of low-energy conduction states
does not increase with film thickness.
%
%
Thus, for a Fermi energy fixed slightly below the band edge,
the electron density and the conductivity 
decrease as the number of layers increase as shown in Tables 
\ref{tab:GaS_thermo} - \ref{tab:InSe_thermo}.
As a result, the maximum n-type ZT for each material
occurs at a single monolayer and decreases 
with each additional layer. 
\begin{centering}
\vspace{0.2in}
\begin{table*}
\begin{tabular}{l c c c c | c c c c}
\hline\hline
Thickness & \multicolumn{1}{c}{p} & \multicolumn{1}{c}{$S_{p}$} &  \multicolumn{1}{c}{$ \sigma_{p}$}  &  \multicolumn{1}{c}{ZT$_{p}$} &\multicolumn{1}{c}{n} & \multicolumn{1}{c}{$\mid$|S$_{e}\mid$} &  \multicolumn{1}{c}{$\sigma_{e}$}  &  \multicolumn{1}{c}{ZT$_{e}$} \\[0.5ex]  
  & \multicolumn{1}{c}{($\times$10$^{18}$cm$^{-3}$)} & \multicolumn{1}{c}{($\mu VK^{-1}$)} &  \multicolumn{1}{c}{($\times 10^{6}\Omega m)^{-1}$}  & \multicolumn{1}{c}{ } & \multicolumn{1}{c}{($\times$10$^{18}$ cm$^{-3}$)} & \multicolumn{1}{c}{($\mu VK^{-1}$)} &  \multicolumn{1}{c}{($\times 10^{6}\Omega m)^{-1}$}   &  \multicolumn{1}{c}{ } \\[0.5ex] 
\hline
1L  & 9.30   & 244.2 & 1.26 &  2.43 & 3.75 & 210.8 &  .210 & .350 \\ [0.5ex] \hline

2L  & 4.20 & 228.7 & .610 &  1.12& 1.63 & 200.0 &  .113 & .181 \\ [0.5ex] \hline

3L  & 2.32 & 229.5 & .361 &  .701 & 1.25 & 196.9 &  .078 & .120\\ [0.5ex] \hline

4L  & 1.91 & 222.0 & .292 &  .532 & 1.02  & 198.1 & .059 & .094 \\ [0.5ex] \hline

Bulk  & 1.30 & 195.1 & .180 & .280 & 1.21 & 179.8 & .070 & .092 \\  [0.5ex] \hline 
\hline 
\end{tabular}
\caption{ 
InS thermoelectric properties for bulk and one to four monolayers at 
$T=300$ K.
Hole and electron carrier concentrations (p and n), Seebeck coefficients (S$_{p}$ and S$_{e}$), 
and electrical conductivties ($\sigma_{p}$ and $\sigma_{n}$) at
the peak p-type and n-type ZT.
}
\label{tab:InS_thermo}
\end{table*}
%
\end{centering}
\begin{centering}
\vspace{0.2in}
\begin{table*}
\begin{tabular}{l c c c c | c c c c}
\hline\hline
Thickness & \multicolumn{1}{c}{p} & \multicolumn{1}{c}{$S_{p}$} &  \multicolumn{1}{c}{$ \sigma_{p}$}  &  \multicolumn{1}{c}{ZT$_{p}$} &\multicolumn{1}{c}{n} &\multicolumn{1}{c}{$\mid$|S$_{e}\mid$} &  \multicolumn{1}{c}{$\sigma_{e}$}  &  \multicolumn{1}{c}{ZT$_{e}$} \\[0.5ex]  
  & \multicolumn{1}{c}{($\times$10$^{18}$cm$^{-3}$)} & \multicolumn{1}{c}{($\mu VK^{-1}$)} &  \multicolumn{1}{c}{($\times 10^{6}\Omega m)^{-1}$}  & \multicolumn{1}{c}{ } & \multicolumn{1}{c}{($\times$10$^{18}$ cm$^{-3}$)} & \multicolumn{1}{c}{($\mu VK^{-1}$)} &  \multicolumn{1}{c}{($\times 10^{6}\Omega m)^{-1}$}   &  \multicolumn{1}{c}{ } \\[0.5ex] 
\hline
1L    & 9.71 & 229.8 & .981 &  1.08 & 2.34 & 200.5 &  .192 & .180 \\ [0.5ex] \hline

2L  & 4.04  & 219.8 & .430 & .471 &  1.22 &  194.7 & .111 &  .090 \\ [0.5ex] \hline

3L    & 4.18 & 204.2 & .471 & .292 & .781 & 189.1 &  .067 & .059\\ [0.5ex] \hline

4L    & 2.45 & 201.0 & .261  &  .252 & .610 & 186.8 &  .053 & .045 \\ [0.5ex] \hline

Bulk  &  1.75 & 179.1 & .181 & .142 & .652 & 160.9 &  .054 & .034 \\ [0.5ex] \hline
\hline 
\end{tabular}
\caption{ 
InSe
thermoelectric properties for bulk and one to four monolayers at 
$T=300$ K.
Hole and electron carrier concentrations (p and n), Seebeck coefficients (S$_{p}$ and S$_{e}$), 
and electrical conductivties ($\sigma_{p}$ and $\sigma_{n}$) at
the peak p-type and n-type ZT.
}
\label{tab:InSe_thermo}
\end{table*}
%
\end{centering}

\subsection{Bi$_{2}$Se$_{3}$}
\label{sec:Bi2Se3}
Bi$_{2}$Se$_{3}$ is an iso-structural compound of the 
well known thermoelectric, Bi$_{2}$Te$_{3}$.
Both materials have been intensely studied recently 
because they are also topological insulators.\cite{Bi2Te3_bulkARPES_Science09,
GYin_TI_JAP13,Hasan:BiSe:Nature:2008}
Bulk Bi$_{2}$Se$_{3}$ has been studied less for its thermoelectric 
properties due to its slightly
higher thermal conductivity compared to Bi$_{2}$Te$_{3}$.
The bulk thermal conductivity of Bi$_{2}$Se$_{3}$ is 
2 W-(mK)$^{-1}$ compared to a bulk thermal conductivity of 
1.5 W-(mK)$^{-1}$ reported for Bi$_{2}$Te$_{3}$.
\cite{goldsmid2009thermo,goyal_Balandin}
However, the thermoelectric performance of bulk Bi$_{2}$Te$_{3}$ is limited
to a narrow temperature window around room temperature 
because of its small
bulk band gap of approximately 160 meV.\cite{Bi2Te3_bulkARPES_Science09}
The band gap of single quintuple layer (QL) Bi$_{2}$Te$_{3}$ was previously
calculated to be 190 meV.\cite{Zahid_Lake}
In contrast, the bulk bandgap of Bi$_{2}$Se$_{3}$ is $\sim$300 meV \cite{Bi2Se3_bulkgap_Cava_PRL}
which allows it to be utilized at higher temperatures.
\begin{figure}[!h]
\includegraphics[width=5in]{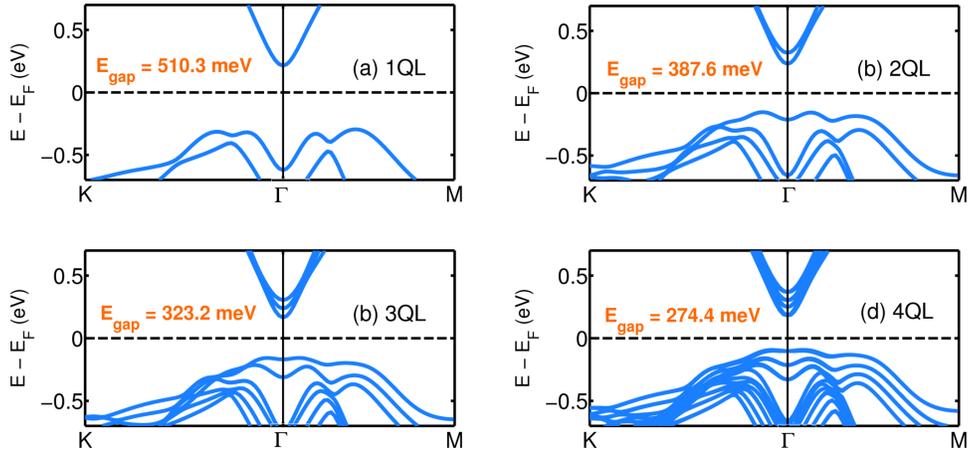}
\caption{
(Color online) Ab-initio band structure including spin-orbit
interaction of Bi$_{2}$Se$_{3}$: (a) 1 QL, (b) 2 QL, (c) 3 QL and (d) 4 QL.  
}
\label{fig:Ek_Bi2Se3}
\end{figure}	

The optimized lattice parameters for bulk Bi$_{2}$Se$_{3}$ 
are listed in Table \ref{tab:mat_params}.
The optimized bulk crystal structure and bulk band gap is consistent with prior
experimental and theoretical studies of bulk Bi$_{2}$Se$_{3}$.
\cite{nakajima_Bi2Se3_structure_expt,Udo_Bi2Se3}
Using the optimized lattice parameters of the bulk structure, 
the electronic
structures of one to four quintuple layers of Bi$_{2}$Se$_{3}$ are calculated
with the inclusion of spin-orbit coupling.
The electronic structures of 1 to 4 QLs of 
Bi$_{2}$Se$_{3}$ are shown in 
Figure \ref{fig:Ek_Bi2Se3}.
The band gaps for one to four quintuple layers of Bi$_{2}$Se$_{3}$
are 510 meV, 388 meV, 323 meV and 274 meV for the 1QL, 2QL, 3QL and 4QL
films, respectively.
The effective masses of the conduction and valence band at $\Gamma$ for 1QL to 4QL of 
Bi$_{2}$Se$_{3}$ are listed in Table \ref{tab:bi2se3_mass}.

For each of the thin film structures, the 
conduction bands are parabolic and located at $\Gamma$. 
The conduction band at $\Gamma$ of the 1QL structure is composed of 
13$\%$ Se $s$, 24$\%$ Se $p_{xy}$,
16$\%$ Bi $p_{xy}$, and 39$\%$ Bi $p_{z}$.
The orbital composition of the $\Gamma$ valley remains qualitatively
the same as the film thickness increases to 4QL.
The orbital composition of the bulk conduction band 
is 79$\%$ Se $p_{z}$ and 16$\%$ Bi $s$.
As the film thickness increases above 1QL, the conduction band at $\Gamma$
splits, as illustrated in Figs. \ref{fig:Ek_Bi2Se3}(b)-(d).
In the 2QL, 3QL and 4QL structures the conduction band splitting 
varies between 53.9 meV and 88.2 meV.
As with the III-VIs, the number of low-energy 
conduction band states per unit thickness
decreases with increasing thickness.

\begin{figure}[!h]
\includegraphics[width=5in]{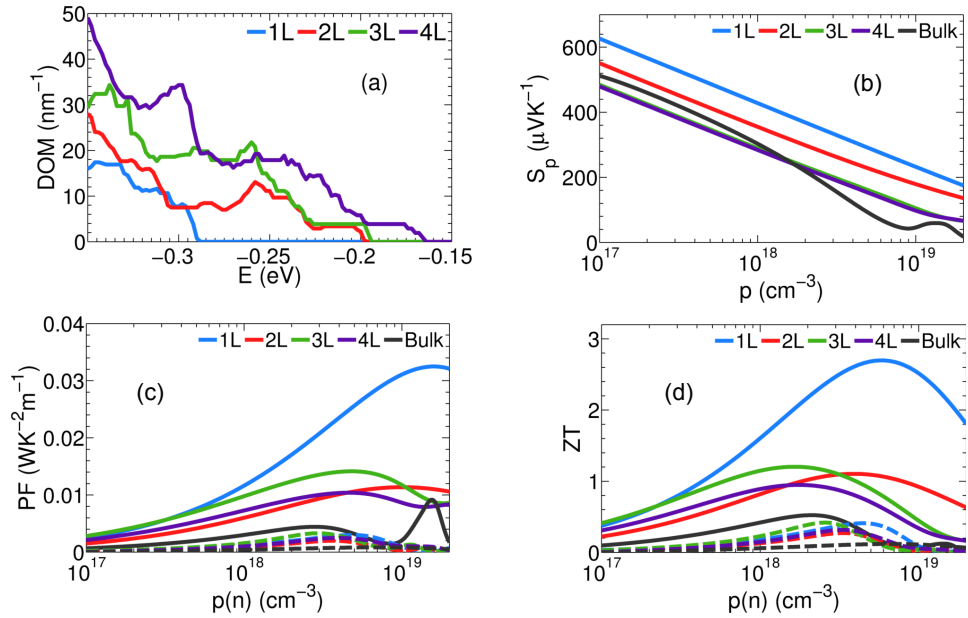}
\caption{(Color online) (a) Distribution of modes per unit width versus energy
 for Bi$_{2}$Se$_{3}$. The midgap energy is set to E=0. 
Thermoelectric properties of p-type (solid line)
and n-type (broken line) Bi$_{2}$Se$_{3}$: (b) Seebeck coefficient, 
(c) power factor and (d) thermoelectric figure-of-merit, ZT, at room temperature 
for 1L (blue), 2L (red), 3L (green), 4L (purple) and bulk (black) }
\label{fig:Bi2Se3_thermo}
\end{figure}

The valence bands have slightly anistropic Mexican hat dispersions.
The values of $\epsilon_{0}$ and $k_0$ used to characterize the Mexican hat for the 1QL to 4QL 
structures of Bi$_{2}$Se$_{3}$ are listed in Table \ref{tab:e0_k0_III_VIs}.
The radius k$_{0}$ is the distance from $\Gamma_{v}$ to the band extremum
along $\Gamma_{v} - M_{v}$, which is the valence band maxima 
for the 1QL to 4QL structures.
The energy difference between the valence band maxima and the band extremum 
along $\Gamma_{v} - K_{v}$ decreases from 19.2 meV to 0.56 meV
as the film thickness increases from 1QL to 4QL.
The Mexican hat dispersion in 1QL of Bi$_{2}$Se$_{3}$
is better described as a double brimmed hat consisting of two concentric rings
in $k$-space
characterized by four points of extrema that are nearly degenerate.
The band extremum along $\Gamma_{v} - M_{v}$ adjacent to the
valence band maxima, is 36 meV below the valence band maxima.
Along $\Gamma_{v} - K_{v}$ the energy difference between 
the two band extrema is 4.2 meV.
At $\Gamma_{v}$, the orbital composition of the valence band for 
1QL of Bi$_{2}$Se$_{3}$ is 
63$\%$ p$_{z}$ orbitals of Se, 11$\%$ p$_{xy}$ orbitals of Se
and 18$\%$ s orbitals of Bi,
and the orbital composition remains qualitatively the same as the film thickness increases
to 4QL.
As the thickness increases above a monolayer, the energy splitting of the valence
bands from each layer is large with respect to room temperature $k_BT$
and more complex than the splitting seen in the III-VIs. 
At a bilayer, the highest valence band loses most of the outer $k$-space ring,
the radius $k_0$ decreases by a factor 3.1
and the height ($\ep_0$) of the hat decreases by a factor of 5.1.
This decrease translates into a decrease in the initial step height of the
density of modes shown in Figure \ref{fig:Bi2Se3_thermo}(a). 
The second highest valence band retains most of the shape of the original monolayer
valence band, but it is now too far from the valence band edge to contribute
to the low-energy electronic or thermoelectric properties.
Thus, Bi$_2$Se$_3$ follows the same trends as seen in Bi$_2$Te$_3$;
the large enhancement in the thermoelectric properties resulting
from bandstructure are only significant for a monolayer \cite{Lundstrom_Jesse_Bi2Te3}.
\begin{table}[!h]
\begin{tabular} {c | c | c }
\hline\hline
Structure &  $\Gamma_{v}$ (m$_{0}$)  & $\Gamma_{c}$ (m$_{0}$)  \\[0.5ex]  \hline
1L & 0.128 & 0.132 \\ [0.5ex]
2L & 0.436 & 0.115 \\ [0.5ex]
3L & 1.435 & 0.176 \\ [0.5ex]
4L & 1.853 & 0.126 \\ [0.5ex]
\hline
\end{tabular}
\caption{Ab-initio calculations of the hole and electron effective masses
at the $\Gamma$-valley valence and conduction band edges of Bi$_2$Se$_2$. }
\label{tab:bi2se3_mass} 
\end{table}

\begin{centering}
\begin{table*}
\begin{tabular}{l c c c c | c c c c}
\hline\hline
Thickness & \multicolumn{1}{c}{p} & \multicolumn{1}{c}{$S_{p}$} &  \multicolumn{1}{c}{$ \sigma_{p}$}  &  \multicolumn{1}{c}{ZT$_{p}$} &\multicolumn{1}{c}{n} & \multicolumn{1}{c}{$\mid$|S$_{e}\mid$} &  \multicolumn{1}{c}{$\sigma_{e}$}  &  \multicolumn{1}{c}{ZT$_{e}$} \\[0.5ex]  
  &  \multicolumn{1}{c}{($\times$10$^{18}$cm$^{-3}$)} & \multicolumn{1}{c}{($\mu VK^{-1}$)} &  \multicolumn{1}{c}{($\times 10^{6}\Omega m)^{-1}$}  & \multicolumn{1}{c}{ } & \multicolumn{1}{c}{($\times$10$^{18}$ cm$^{-3}$)} & \multicolumn{1}{c}{($\mu VK^{-1}$)} &  \multicolumn{1}{c}{($\times 10^{6}\Omega m)^{-1}$}   &  \multicolumn{1}{c}{ } \\[0.5ex] 
\hline
1L   & 7.66 & 279.3 & .371 &  2.86 & 4.63 & 210.1 &  .067 & .411 \\ [0.5ex] \hline

2L   & 4.65 & 251.3 & .282 &  1.17 & 3.38 & 208.2 &  .049 & .271 \\ [0.5ex] \hline

3L   & 2.77 & 259.4 & .172 &  1.12 & 2.96 & 198.3 &  .043 & .232 \\ [0.5ex] \hline

4L   & 2.58 & 237.8 & .161 &  .942 & 2.56 & 185.8 &  .037 & .190 \\ [0.5ex] \hline

Bulk  & 1.95 & 210.7 & .095 & .521 & 1.23 & 191.9 &  .020 & .123 \\ [0.5ex] \hline

\hline 
\end{tabular}
\caption{ 
Bi$_{2}$Se$_{3}$
thermoelectric properties for bulk and one to four quintuple layers at 
$T=300$ K.
Hole and electron carrier concentrations (p and n), Seebeck coefficients (S$_{p}$ and S$_{e}$), 
and electrical conductivties ($\sigma_{p}$ and $\sigma_{n}$) at
the peak p-type and n-type ZT.
}
\label{tab:Bi2Se3_thermo}
\end{table*}
%
\end{centering}

The p-type and n-type Seebeck coefficient, electrical conductivity,
power factor and the thermoelectric figure-of-merit (ZT) 
as a function of carrier concentration at room temperature
for Bi$_{2}$Se$_{3}$ are illustrated in Figure \ref{fig:Bi2Se3_thermo}.
The thermoelectric parameters at $T=300$ K
of bulk and one to four quintuple layers
for Bi$_{2}$Se$_{3}$ are summarized in Table \ref{tab:Bi2Se3_thermo}.

The p-type ZT for the single quintuple layer is enhanced by a factor
of 5.5 compared to that of the bulk film.
At the peak ZT, the hole concentration
is 4 times larger than that of the bulk,
and the position of the Fermi energy with
respect to the valence band edge ($E_F - E_V$) is 
45 meV higher than that of the bulk.
The bulk and monolayer magnitudes of $I_0$ ($I_1$) are
0.88 (2.14) and 3.45 (11.2), respectively, 
giving increases of 3.9 (5.2) as the thickness is reduced from bulk to monolayer.
As the film thickness is reduced from 4 QL to 1 QL, 
the magnitudes of $I_{0}$ and $I_{1}$ at the peak ZT 
increase by factors of 2.4 and 2.8, respectively.

The peak room temperature n-type ZT also occurs for 1QL of Bi$_{2}$Se$_{3}$.
In one to four
quintuple layers of Bi$_{2}$Se$_{3}$, two degenerate bands at 
$\Gamma$ contribute to the conduction band density of modes.
The higher $\Gamma$ valleys contribute little to the
conductivity as the film thickness increases.
The Fermi levels 
at the peak n-type, room-temperature ZT 
rise from 34 meV to 12 meV
below the conduction band edge as the film
thickness increases from 1 QL to 4 QL while the
electron density decreases by a factor of 1.8.
This results in a maximum n-type ZT for the 1QL structure.

A recent study on the thickness dependence of the thermoelectric properties
of ultra-thin Bi$_{2}$Se$_{3}$ obtained a p-type ZT value of 0.27 and a 
p-type peak power factor of 0.432 mWm$^{-1}$K$^{-2}$ for the 1QL film.
\cite{Udo_Bi2Se3}
The differences in the power factor and the ZT are due to the different
approximations made in the relaxation time (2.7 fs)
and lattice thermal conductivity (0.49 W/mK)
used in this study.
Using the parameters of Ref.[\onlinecite{Udo_Bi2Se3}] in our density of modes
calculation of 1QL of Bi$_{2}$Se$_{3}$ gives a 
peak p-type ZT of 0.58 and peak p-type power factor
of 0.302 mWm$^{-1}$K$^{-2}$.
We also compare the thermoelectric properties of single quintuple layer
Bi$_{2}$Se$_{3}$ and Bi$_{2}$Te$_{3}$.
In both materials, the valence band of the single quintuple film
is strongly deformed into a Mexican hat.
The radius $k_{0}$ for 1QL of Bi$_{2}$Se$_{3}$ is a factor of $\sim$2 
higher than $k_{0}$ for 1QL Bi$_{2}$Te$_{3}$.
The peak p-type ZT of 7.15 calculated for Bi$_{2}$Te$_{3}$
\cite{Zahid_Lake} is a factor of 2.5
higher than the peak p-type ZT of 2.86 obtained for a single quintuple layer of
Bi$_{2}$Se$_{3}$.
This difference in the thermoelectric figure of merit can
be attributed to the different
approximations in the hole mean free path chosen 
for Bi$_{2}$Se$_{3}$ ($\lambda_{p}$=2.4 nm)
and Bi$_{2}$Te$_{3}$ ($\lambda_{p}$=8 nm) \cite{Zahid_Lake} and
the higher lattice thermal conductivity of 
 Bi$_{2}$Se$_{3}$ ($\kappa_{l}$=2 W/mK)
compared to Bi$_{2}$Te$_{3}$ ($\kappa_{l}$=1.5 W/mK).

\subsection{Bilayer Graphene}
\label{sec:BiGraphene}
AB stacked bilayer graphene (BLG) is a gapless semiconductor
with parabolic conduction and valence bands that are located
at the $K$ ($K'$) symmetry points.
Prior experimental \cite{Wang_exp_bi_gap_Nat09,Falko_BLG_Lifshitz_PRL14} and 
theoretical \cite{MacDonald_bi_gap_PRB07} studies demonstrated the formation of 
a bandgap in BLG with the application of a vertical electric field.
The vertical electric field also deforms the conduction
and valence band edges at $K$ into a Mexican-hat 
dispersion \cite{Fermi_ring_Neto_PRB07,Falko_BLG_Lifshitz_PRL14}.
Using ab-initio calculations we compute the band structure of bilayer graphene subject
to vertical electric fields ranging from 0.05 V/\AA~to 0.5 V/\AA.
The lattice parameters for the bilayer graphene structure used in our simulation
are given in Table \ref{tab:mat_params}.
The ab-initio calculated band gaps are in good agreement with prior
calculations. \cite{MacDonald_bi_gap_PRB07, McCann_Eg_bilayer_PRB06}
The bandgap increases from 144.4 meV to 277.3 meV as the applied field increases
from 0.05 V/\AA~to 0.5 V/\AA.

\begin{figure}[!h]
\includegraphics[width=5in]{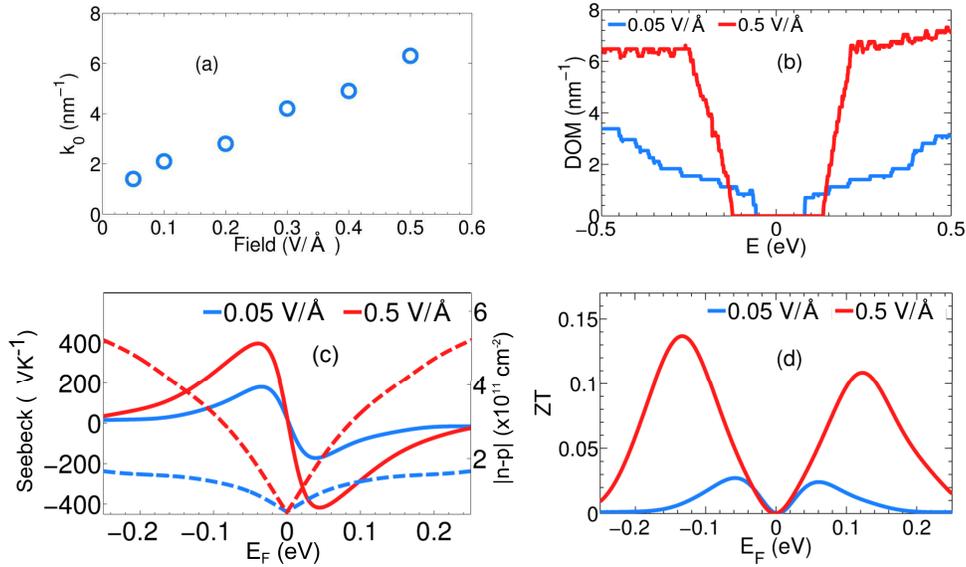}
 \caption{(Color online) (a) Evolution of the radius of the Mexican hat, k$_{0}$ in
bilayer graphene as a function of an applied vertical electric field.  
(b) Density of modes per unit width for two different vertical fields of 
0.05 V/\AA~(blue) and 0.5 V/\AA~(red).
(c) Seebeck coefficients (solid lines) and carrier concentrations (broken lines) 
for two different vertical fields. 
(d) ZT of bilayer graphene as a function of the Fermi level for two different
vertical fields.
}
\label{fig:Bilayergraphene_Ek}
\end{figure}

For each applied field ranging from from 0.05 V/\AA~to 0.5 V/\AA~both the 
valence band and the conduction band edges lie along the path $\Gamma - K$,
and the radius $k_{0}$ is the distance from $K$
to the band edge along $\Gamma - K$.
The magnitude of $k_{0}$ increases linearly with
the electric field as shown in Figure \ref{fig:Bilayergraphene_Ek}(a).
The dispersions of the valence band and the conduction band quantitatively differ,
and $k_0$ of the valence band is up to 20$\%$ higher than $k_0$ of the
conduction band.
The anisotropy of the conduction and valence Mexican hat dispersions
increase with increasing vertical field.
The extremum point along $K - M$ of the valence (conduction) band
Mexican hat dispersion is lower (higher) in energy 
compared to the band extremum along $\Gamma - K$.
As the field increases from 0.05 V/\AA~to 0.5 V/\AA~the 
energy difference between the two extrema points
increases from 5.2 meV to 69.4 meV in the valence band and
7.7 meV to 112.3 meV in the conduction band.
This anisotropy in the Mexican hat of the valence and conduction
band leads to a finite slope in the density of modes illustrated
in Figure \ref{fig:Bilayergraphene_Ek}(b).

As the applied field is increased from 0.05 V/\AA~to 0.5 V/\AA~the height of the density
of modes step function in the valence and conduction band increases by a factor of 
5.7.
Figure \ref{fig:Bilayergraphene_Ek}(b) illustrates the density of modes distribution
for the conduction and valence band states for the lowest field applied (0.05 V/\AA)
and the highest field applied (0.5 V/\AA).
The p-type thermoelectric parameters 
of bilayer graphene subject to vertical electric fields ranging 
from 0.05 V/\AA~to 0.5 V/\AA~are summarized in Table \ref{tab:BiGraphene_thermo}.
The p-type and n-type thermoelectric parameters are similar. 
Figure \ref{fig:Bilayergraphene_Ek}(d) compares the calculated ZT versus Fermi level
for bilayer graphene
at applied electric fields of 0.05 V/\AA~to 0.5 V/\AA.
For an applied electric field of 0.5 V/\AA~the p-type and n-type ZT is 
enhanced by a factor of 6 and 4 
in bilayer graphene compared to the ZT of bilayer graphene with no applied electric field.
\begin{centering}
\begin{table*}
\begin{tabular}{l c c c c}
\hline\hline
Field  & p  & $S_{p}$ &  $\sigma_{p}$ & ZT$_{p}$ \\[0.5ex] 
 (V/\AA) &  ($\times 10^{12}$ cm$^{-2}$) & $(\mu VK^{-1})$ &   ($\times 10^{7}\Omega m)^{-1}$) &  \\[0.5ex] 
\hline
0.0  & .12 & 138.4 &  .83  & .0230  \\ [0.5ex] \hline

0.05  & .11  & 154.9  &  .77  &  .0270   \\ [0.5ex] \hline

0.1  & .16 & 192.1  &  1.1  &  .0281   \\ [0.5ex] \hline

0.2   & .19 & 190.7 &  1.3  &  .0693  \\ [0.5ex] \hline

0.3   & .21 & 179.8 &  1.4  &  .0651  \\ [0.5ex] \hline

0.4   & .27  & 196.4 &  1.8  &  .1001    \\ [0.5ex] \hline

0.5   & .31  & 188.0 &  2.1  &  .1401   \\ [0.5ex] \hline
\hline 
\end{tabular}
\caption{Bilayer graphene p-type thermoelectric properties as a function of vertical electric
field at $T=300$ K.
Hole carrier concentrations, p-type Seebeck coefficient, and
electrical conductivity at the peak p-type ZT.}
\label{tab:BiGraphene_thermo}
\end{table*}
%
\end{centering}

\subsection{Bi Monolayer}
\label{sec:Bi}
\begin{figure}[!h]
\includegraphics[width=5in]{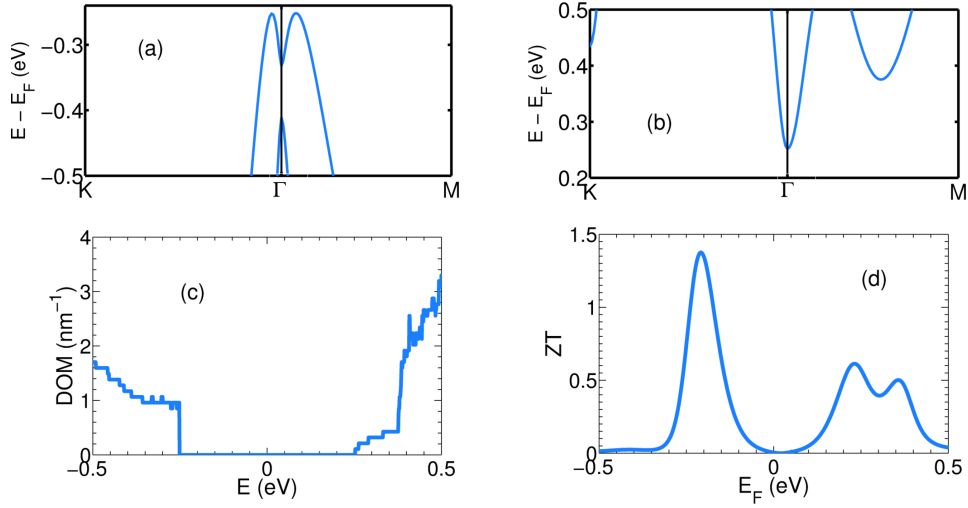}
\caption{(Color online) Electronic structure and thermoelectric properties
of Bi(111) monolayer.  (a) Valence band, (b) Conduction band of Bi(111) monolayer
with spin-orbit interaction. (c) Density of modes with SOC interactions included, (c) Thermoelectric
figure of merit, ZT, at room temperature.}
\label{fig:Bi_Ek_thermo}
\end{figure}

The large spin-orbit interaction in bismuth leads to a 
Rasha-split dispersion of the valence band in a single monolayer
of bismuth.
The lattice parameters for the Bi(111) monolayer used for the 
SOC ab-initio calculations are 
summarized in Table \ref{tab:mat_params}.
The bandgap of the bismuth monolayer is 503 meV
with the conduction band at $\Gamma_{c}$.
The inclusion of spin-orbit interaction splits the two
degenerate bands at $\Gamma_{v}$ by 79 meV and 
deforms the valence band maxima into a Rashba split band.
The calculated band structure of the Bi(111) monolayer is shown in
Figure \ref{fig:Bi_Ek_thermo}(a,b).
The Rashba parameter for the bismuth monolayer
is extracted from the ab-initio calculated
band structure.
The curvature of the 
valence band maxima of the Rashba band gives an effective
mass of $m^* = 0.1351$.
The vertical splitting of the bands at small $k$
gives an $\alpha_{R} = 2.142$ eV\AA .
Prior experimental and theoretical studies on the strength
of the Rashba interaction in Bi(111) surfaces
demonstrate $\alpha_R$ values ranging from 
0.55 eV\AA$^{-1}$ to 3.05 eV\AA$^{-1}$ .
\cite{BiTeI_Rashba_NatMat11}
A slight asymmetry in the Rashba-split dispersion leads
to the valence band maxima lying along $\Gamma_{v} - M_{v}$.
The band extremum along $\Gamma_{v} - K_{v}$ is 0.5 meV below 
the valence band maxima.
The radius of the valence band-edge $k_0$, which is the distance
from $\Gamma_{v}$ to the band extremum along $\Gamma_{v} - M_{v}$
is 1.40 nm$^{-1}$ similar to 4L InSe.
The valence band-edge density of modes shown in Fig. \ref{fig:Bi_Ek_thermo}(c) 
is a step function with a peak height of 0.96 nm$^{-1}$.
Figure \ref{fig:Bi_Ek_thermo}(d) shows
the resulting thermoelectric
figure of merit ZT as a function of Fermi level position
at room temperature.
The thermoelectric parameters at $T=300$ K 
are summarized in Table \ref{tab:Bi_thermo}.

\begin{table}[!h]
\begin{tabular}{ >{\centering}p{1.1cm} p{0.9cm} p{1.1cm} p{0.9cm} |>{\centering} p{1.1cm} p{0.9cm} p{1.1cm} p{0.9cm}}
\hline\hline
\multicolumn{1}{c}{p} & \multicolumn{1}{c}{$S_{p}$} &  \multicolumn{1}{c}{$ \sigma_{p}$}  &  \multicolumn{1}{c}{ZT$_{p}$} &\multicolumn{1}{c}{n} &\multicolumn{1}{c}{$\mid$|S$_{e}\mid$} &  \multicolumn{1}{c}{$\sigma_{e}$}  &  \multicolumn{1}{c}{ZT$_{e}$} \\[0.5ex]  
 \multicolumn{1}{c}{($\times$10$^{19}$cm$^{-3}$)} & \multicolumn{1}{c}{($\mu VK^{-1}$)} &  \multicolumn{1}{c}{($\times 10^{6}\Omega m)^{-1}$}  & \multicolumn{1}{c}{ } & \multicolumn{1}{c}{($\times$10$^{19}$ cm$^{-3}$)} & \multicolumn{1}{c}{($\mu VK^{-1}$)} &  \multicolumn{1}{c}{($\times 10^{6}\Omega m)^{-1}$}   &  \multicolumn{1}{c}{ } \\[0.5ex] 
\hline
 .61 & 239.7 & .39 &  1.38 & .35 & 234.1 &  .19 & .61 \\ [0.5ex] \hline
\hline 
\end{tabular}
\caption{ 
Bi(111) thermoelectric properties at $T=300$ K.
Hole and electron carrier concentrations (p and n), Seebeck coefficients (S$_{p}$ and S$_{e}$), 
and electrical conductivties ($\sigma_{p}$ and $\sigma_{n}$)
at the peak p-type and n-type ZT.
}
\label{tab:Bi_thermo}
\end{table}

Using mean free paths of $\lambda_{e}$=50nm for electrons and $\lambda_{p}$=20nm
for holes, our peak ZT values are consistent with
a prior report on the thermoelectric properties of monolayer
Bi.\cite{Bi_1L_thermo_JPC}.
The peak p-type (n-type) ZT and Seebeck values of 2.3 (1.9) and
786 $\mu$V/K (-710 $\mu$V/K) are consistent with reported values of
2.4 (2.1) and 800 $\mu$V/K (-780 $\mu$V/K) in
Ref.[\onlinecite{Bi_1L_thermo_JPC}].

\section{Summary and Conclusions}
Monolayer and few-layer structures of III-VI materials (GaS, GaSe, InS, InSe), 
Bi$_{2}$Se$_{3}$, monolayer Bi, and biased 
bilayer graphene all have a valence band that forms a ring in 
$k$-space.
For monolayer Bi, the ring results from Rashba splitting of the spins.
All of the other few-layer materials have valence bands in the shape of
a `Mexican hat.'
For both cases, a band-edge that forms a ring in $k$-space
is highly degenerate.
It coincides with a singularity in the density of states
and a near step-function turn-on of the density of modes at
the band edge.
The height of the step function is approximately proportional to the radius
of the $k$-space ring.

The Mexican hat dispersion in the valence band of the 
III-VI materials exists for few-layer geometries, and it is most
prominent for monolayers, which have the largest radius $k_0$
and the largest height $\ep_0$.  
The existence of the Mexican hat dispersions 
and their qualitative features
do not depend on the choice of functional, stacking, or the inclusion or omission
of spin-orbit coupling, and recent calculations by others show that they 
are also unaffected by many-electron self-energy effects.\cite{SGLouie_GaSe_arxiv}
At a thickness of 8 layers, all of the III-VI valence band dispersions are parabolic.

The Mexican hat dispersion in the valence band of monolayer Bi$_{2}$Se$_{3}$
is qualitatively different from those in the monolayer III-VIs.
It can be better described as a double-brimmed hat characterized by 4 points of extrema that
lie within $\sim k_BT$ of each other at room temperature.
Futhermore, when two layers are brought together to form a bilayer, the 
energy splitting of the two valence bands in each layer
causes the highest band to lose most of its outer ring causing a large decrease
in the density of modes and reduction in the thermoelectric properties.
These trends also apply to Bi$_{2}$Te$_{3}$. \cite{Lundstrom_Jesse_Bi2Te3}

The valence band of monolayer Bi also forms a $k$-space ring that results from
Rashba splitting of the bands. 
The diameter of the ring is relatively small compared to those of monolayer Mexican
hat dispersions.
However, the ring is the most isotropic of all of the monolayer materials considered,
and it gives a very sharp step function to the valence band density of modes.

As the radius of the $k$-space ring increases, the Fermi level
at the maximum power factor or ZT moves higher into the bandgap
away from the valence band edge. 
Nevertheless, the hole concentration increases.
The average energy carried by a hole with respect to the Fermi energy increases.
As a result, the Seebeck coefficient increases.
The dispersion with the largest radius coincides with the maximum power factor provided
that the mean free paths are not too different.
For the materials and parameters considered here, the dispersion with the largest
radius also results in the largest ZT at room temperature.
Bilayer graphene may serve as a test-bed to measure these effects, 
since a cross-plane electric field linearly increases
the diameter of the Mexican hat ring, and 
the features of the Mexican hat in bilayer graphene have recently been 
experimentally observed.\cite{Falko_BLG_Lifshitz_PRL14}

With the exception of monolayer GaS, 
the conduction bands of few-layer
n-type III-VI and Bi$_2$Se$_3$ compounds 
are at $\Gamma$ with a significant $p_z$ orbital component. 
In bilayers and multilayers, these bands couple and split pushing the 
added bands to higher energy above the thermal transport window.
Thus, the number of low-energy states per layer is maximum for a monolayer.
In monolayer GaS, the conduction band is at M with 3-fold valley degeneracy.
At thicknesses greater than a monolayer, the GaS conduction band is at $\Gamma$,
the valley degeneracy is one, and the same splitting of the bands occurs
as described above.
Thus, the number of low-energy states per layer is also maximum for a monolayer GaS.
This results in maximum values for the n-type Seebeck coefficients,
power factors, and ZTs at monolayer thicknesses for all of these materials.

\noindent
\begin{acknowledgements}
This work is supported in part by the National Science Foundation (NSF)
Grant Nos. 1124733 
and the Semiconductor Research Corporation (SRC) Nanoelectronic Research Initiative
as a part of the Nanoelectronics for 2020
and Beyond (NEB-2020) program, FAME, one of six centers of STARnet, a 
Semiconductor Research Corporation program sponsored by MARCO and DARPA.
This work used the Extreme Science and Engineering Discovery Environment (XSEDE), 
which is supported by National Science Foundation grant number OCI-1053575.
\end{acknowledgements}

\end{document}